\documentclass[twocolumn]{aastex7}

\usepackage{amsmath}
\usepackage{soul}
\usepackage{comment}
\newcommand \bx {\boldsymbol{x}}
\defcitealias{2021A&A...647A.186S}{ST21}
\shorttitle{BALLAD-POL V}
\shortauthors{Singh et al.}

\submitjournal{ApJ}

\begin{document}

\title{B-fields And Dust in InterstelLar FiLAments Using Dust POLarization (BALLAD-POL): V. Magnetic Fields and Grain Alignment Mechanisms in Serpens Main Using JCMT/POL-2 Polarization}

\author[orcid=0009-0000-2098-6119,sname='Singh']{Sadhana Singh}
\affiliation{Department of Physics, Maulana Azad National Institute of Technology (MANIT), Bhopal 462003, India}
\email[show]{sadhana.ftp22@gmail.com}  
    
\author[orcid=0000-0003-2017-0982,sname='Hoang']{Thiem Hoang} 
\affiliation{Korea Astronomy and Space Science Institute, 776 Daedeokdae-ro, Yuseong-gu, Daejeon 34055, Republic of Korea}
\affiliation{Department of Astronomy and Space Science, University of Science and Technology, 217 Gajeong-ro, Yuseong-gu, Daejeon, 34113, Republic of Korea}
\email[show]{thiemhoang@kasi.re.kr}

\author[orcid=0000-0002-2808-0888]{Pham Ngoc Diep}
\affiliation{Center for Astrophysics and Science Exploration, Vietnam National Space Center, Vietnam Academy of Science and Technology, 18 Hoang Quoc Viet, Hanoi, Vietnam}
\affiliation{University of Science and Technology of Hanoi, Vietnam Academy of Science and Technology, 18 Hoang Quoc Viet, Hanoi, Vietnam}
\email{pndiep@vnsc.org.vn}

\author[orcid=0000-0002-5913-5554]{Nguyen Bich Ngoc}
\affiliation{Center for Astrophysics and Science Exploration, Vietnam National Space Center, Vietnam Academy of Science and Technology, 18 Hoang Quoc Viet, Hanoi, Vietnam}
\affiliation{Academia Sinica Institute of Astronomy and Astrophysics, 11F of Astro-Math Bldg, 1, Sec. 4, Roosevelt Rd., Taipei 10617, Taiwan}
\email{capi37capi@gmail.com}

\author[orcid=0000-0003-4022-4132]{Woojin Kwon}
\affiliation{Department of Earth Science Education, Seoul National University, 1 Gwanak-ro, Gwanak-gu, Seoul 08826, Republic of Korea}
\affiliation{SNU Astronomy Research Center, Seoul National University, 1 Gwanak-ro, Gwanak-gu, Seoul 08826, Republic of Korea}
\affiliation{The Center for Educational Research, Seoul National University, 1 Gwanak-ro, Gwanak-gu, Seoul 08826, Republic of Korea}
\email{wkwon@snu.ac.kr}


\begin{abstract}
We investigate magnetic fields ($B$-field) and grain alignment mechanisms in Serpens Main using 850 $\mu$m dust polarization observations from JCMT/POL-2. We derive a map of $B$-field strength and calculate the mass-to-flux ratio and Alfv\'enic Mach number, suggesting the subdominance of $B$-fields over gravity but their dominance over turbulence. Fitting the Ricean mean model to the nondebiased polarization fraction ($P$) as a function of emission intensity ($I$) yields a slope $\alpha = 0.68\pm0.01$, revealing   depolarization. To understand its origins, we examine the dependence of $P$ on gas column density ($N({\rm H}_{2})$), dust temperature ($T_{\rm d}$), and polarization angle dispersion function ($S$). We find no correlation between $P$ and $S$, whereas for $P\times S$, an observational proxy for alignment efficiency decreases with increasing $N({\rm H}_{2})$. This suggests that $B$-field tangling is not the primary cause of depolarization and instead points to reduced grain alignment in denser regions. To test this interpretation within radiative torque (RAT) alignment theory, we calculate the minimum size of aligned grains ($a_{\rm align}$) and magnetic relaxation strength ($\delta_{\rm mag,sp}$) using $N({\rm H}_{2})$, $T_{\rm d}$, and our estimated $B$-field map. We find that both $P$ and $P \times S$ decrease with increasing $a_{\rm align}$, supporting reduced grain alignment as the main cause of depolarization in Serpens Main. However, neither $P$ nor $P \times S$ shows a correlation with $\delta_{\rm mag,sp}$, suggesting that  enhanced grain alignment due to magnetic relaxation is insufficient to compensate for the reduced alignment associated with increasing $a_{\rm align}$ toward denser regions.
\end{abstract}


\section{Introduction} \label{sec:intro}
Molecular clouds (MCs) are the nurseries where new stars are formed. Herschel observations \citep{2010A&A...518L...1P} reveal that MCs consist of filamentary structures with dense embedded cores \citep{2010A&A...518L.100M, 2010A&A...518L.102A, 2014prpl.conf...27A, 2023A&A...669A.120Z}. However, how such dense cores form and collapse to form young stars is still hotly debated. Modern understanding posits that star formation is governed by the complex interplay between gravity, turbulence, magnetic fields, and stellar feedback. Strong $B$-fields play a crucial role in supporting MCs against gravitational collapse, channeling turbulent flows, and driving the accretion of matter \citep[see the review by][]{2012ARA&A..50...29C}. Therefore, accurate measurements of $B$-fields in MCs are required for understanding their role relative to gravity and turbulence in star formation.

Polarization of thermal dust emission from aligned, nonspherical dust grains is a powerful tool for probing $B$-fields in the interstellar medium (ISM) and MCs \citep{2017ApJ...846..122P}. However, the underlying physical mechanisms responsible for grain alignment are not yet fully understood. Different mechanisms have been proposed to explain grain alignment, including paramagnetic relaxation \citep{1951ApJ...114..206D}, mechanical alignment \citep{1952MNRAS.112..215G}, and pinwheel mechanisms \citep{1979ApJ...231..404P}. The current leading grain alignment theory is based on radiative torques (RATs) arising from the interaction of anisotropic radiation fields with irregular dust grains \citep[see reviews by][]{2015ARA&A..53..501A,2015psps.book...81L}. The RAT alignment mechanism was first proposed by \cite{1976Ap&SS..43..291D} and later demonstrated numerically by \cite{1996ApJ...470..551D, 1997ApJ...480..633D}. 

The analytical model of RAT alignment theory was introduced by \citet{2007MNRAS.378..910L} and was quantitatively studied in \citet{2008MNRAS.388..117H, 2009ApJ...695.1457H, 2009ApJ...697.1316H,2014MNRAS.438..680H,2015MNRAS.448.1178H}. The classical RAT theory describes that the grain alignment process with $B$-fields includes two stages: (1) internal alignment, in which the grain's shortest axis is aligned along its angular momentum $J$ due to fast internal relaxation, and (2) external alignment, in which the grain angular momentum becomes aligned along the magnetic field due to fast Larmor precession and RATs (e.g., \citealt{hoang2022internal}). 
According to the unified theory of grain alignment (a.k.a. the magnetically enhanced RAT or MRAT mechanism), the RAT alignment is significantly enhanced by superparamagnetic (SPM) relaxation \citep{2008ApJ...676L..25L,2016ApJ...831..159H}, although SPM relaxation alone cannot cause efficient alignment due to thermal rotation. Observations of starlight polarization in various astrophysical environments provide strong empirical support for the RAT alignment theory \citep[see, e.g.,][]{2007ApJ...665..369A, 2008ApJ...674..304W, 2010ApJ...720.1045A, 2011A&A...534A..19A}.

A new physical effect caused by RATs, the so-called radiative torque disruption (RATD), was first discovered by \citet{2019NatAs...3..766H}, and its implications are discussed in \citet{2019ApJ...876...13H}. Physical modeling of dust polarization considering both the grain alignment and rotational disruption by RATs (a.k.a. the RAT paradigm) was first performed by \citet{2020ApJ...896...44L}. They found that the polarization degree decreases with increasing radiation intensity (strength) when RATD occurs because large aligned grains that dominate the polarization are disrupted into smaller fragments. This is the key prediction of the RAT paradigm that needs to be tested with observational data.
 
With advanced facilities equipped with polarimetric capabilities, such as the High-resolution Airborne Wideband Camera Plus of the Stratospheric Observatory for Infrared Astronomy \citep[SOFIA/HAWC+;][]{2018JAI.....740008H, 2018JAI.....740011T}, the James Clerk Maxwell Telescope \citep[JCMT;][]{2013MNRAS.430.2513H, 2016SPIE.9914E..03F}, and the Atacama Large Millimeter/submillimeter Array  \citep[ALMA;][]{2009IEEEP..97.1463W,2014PASJ...66...57A,2015A&A...577A.129B}, it has become possible to probe dense regions of MCs. Under the $B$-ﬁelds In STar forming Region Observations (BISTRO) survey of the JCMT \citep{2017ApJ...842...66W}, star-forming regions (SFRs) are observed to study the role of $B$-fields in the star formation process. Since then, the magnetic fields of numerous dense clouds have been investigated using dust polarization observations \citep[e.g.,][]{2017ApJ...846..122P, 2021ApJ...908...10N, 2022ApJ...941...51H, 2022ApJ...926..163K, 2024ApJ...962..136W}. 
However, only a few studies have been conducted to test the fundamental physics of grain alignment and to probe dust properties within the dense regions of MCs  \citep{2019ApJ...880...27P, 2021ApJ...908...10N, 2021ApJ...907...88P, 2024ApJ...977...32C}.

A significant effort to test the RAT paradigm in SFRs using far-infrared (far-IR) polarization data from SOFIA/HAWC+ was conducted by \citet{2021ApJ...906..115T, 2021ApJ...908..159T, 2021ApJ...923..130T}. However, these studies focus only on two effects, alignment and disruption by RATs; and the effect of magnetic relaxation on RAT alignment (i.e., the MRAT mechanism) was not studied in detail. We note that, in the MRAT mechanism, $B$-field strength is an important parameter that influences the efficiency of grain alignment and disruption \citep{hoang2022internal}. Therefore, a map of $B$-field strength is required to quantify the MRAT mechanism. This motivated the initiation of the BALLAD-POL project - $B$-fields And dust in interstelLar fiLAments using Dust POLarization \citep{2023ApJ...953...66N} to simultaneously investigate the role of $B$-fields in star formation and dust physics using the full polarization data, including the both polarization angle and degree. As part of this BALLAD-POL series \citep{2023ApJ...953...66N,2024ApJ...974..118N,2025ApJ...981..128P,2025ApJ...990...40P,2026arXiv260301899P}, here we focus on the Serpens Main molecular cloud.

Serpens Main is an active star-forming region \citep{2008hsf2.book..693E}, located at a distance of 415$\pm$25 pc \citep{2010ApJ...718..610D}. Gas and dust are primarily concentrated in the northwest (NW) and southeast (SE) subclusters of the cloud \citep{2014ApJ...797...76L, 2002A&A...392.1053O}. \citet{2014ApJ...797...76L} identified the filaments in Serpens Main by inspecting the morphologies in the integrated-intensity map and the velocity-channel maps of the isolated hyperfine component of $\rm N_{2}H^{+} (1-0)$. More prominent filamentary structures and a complicated velocity structure were found in the SE subcluster compared to the NW subcluster. Six filaments named FS1, FS23 (FS2 and FS3 combined), FC1, FC2, and FN1 were identified in the cloud region (see Figure \ref{fig:stokesI}). A characteristic depth of 0.15 pc was found for the cloud, and the length of the filaments was found to be in the range of 0.17-0.33 pc with an average length of $\sim$0.22 pc and widths in the range of 0.03-0.05 pc (average $\sim$0.03 pc). Many submillimeter sources were found to be associated with the Serpens core \citep{1993A&A...275..195C, 1999MNRAS.309..141D}, and several young stellar objects (YSOs) were identified in Serpens Main \citep{2004A&A...421..623K, 2006ApJ...644..307H, 2009ApJS..181..321E, 2009ApJ...691..672O}.  

As part of the BISTRO survey, \citet{2022ApJ...926..163K} used 850 $\mu$m dust polarization observations from JCMT/POL-2 to perform a detailed study on the variation of $B$-field morphology with the gas density structure. They found that the orientation of the $B$-fields is parallel to the less-dense filaments but becomes perpendicular in denser filaments. Further, they found the density regimes that indicate core formation and dragged-in $B$-fields. 
Using the Davis-Chandrasekhar-Fermi (DCF) technique \citep{1951PhRv...81..890D, 1953ApJ...118..116C}, they also estimated the average $B$-field strengths of the individual filaments, which range from 60 to 300 $\mu$G. 
To better understand the relative role of $B$-fields compared with gravity and turbulence in the star formation process, it is crucial to derive a map of $B$-field strength. Such a $B$-field map is necessary to quantify the effect of magnetic relaxation on the RAT alignment.

The paper is organized as follows. The observation and data analysis are presented in Section \ref{sec:data}. Estimates of the $B$-field and its relative role in gravity and turbulence are presented in Section \ref{sec:B_field}. In Section \ref{sec:dust_phy}, we study the grain alignment mechanisms in detail by performing various data analyses. The discussion is given in Section \ref{sec:diss}, followed by the conclusions in Section \ref{sec:conclusion}.

\section{Observational Data: Thermal Dust Polarization}   \label{sec:data}
We use 850 $\mu$m thermal dust polarization data of Serpens Main observed with the JCMT/POL-2 by \citet{2022ApJ...926..163K}. Polarization observations were carried out by inserting the linear polarimeter ({\it POL-2}) into the Submillimeter Common-User Bolometer Array (SCUBA-2) \citep{2013MNRAS.430.2513H} camera of the JCMT. The technical details of the observation are described in \citet{2022ApJ...926..163K}. 

The entire Serpens Main cloud was observed in two pointings in 2016-2017. The observations of the two fields (SE and NW) were performed with pointings centered at R.A.: 18$^{\rm h}$29$^{m}$59$^{s}$.370, decl.: +01\degr12\arcmin52\arcsec.00 and R.A.: 18$^{\rm h}$29$^{m}$49$^{s}$.530, decl.: +01\degr16\arcmin11\arcsec.10; where the R.A. and decl. coordinates are in J2000. The data reduction procedure used to generate the Stokes parameter maps ($I$, $Q$, and $U$) is described in \citet{2022ApJ...926..163K}. The final mosaicked maps of Stokes $I$, $Q$, and $U$ were produced, which have a pixel size of 4\arcsec\, and were calibrated in janskys per beam (Jy/beam). 

\begin{figure*}
    \centering
    \includegraphics[width=\textwidth]{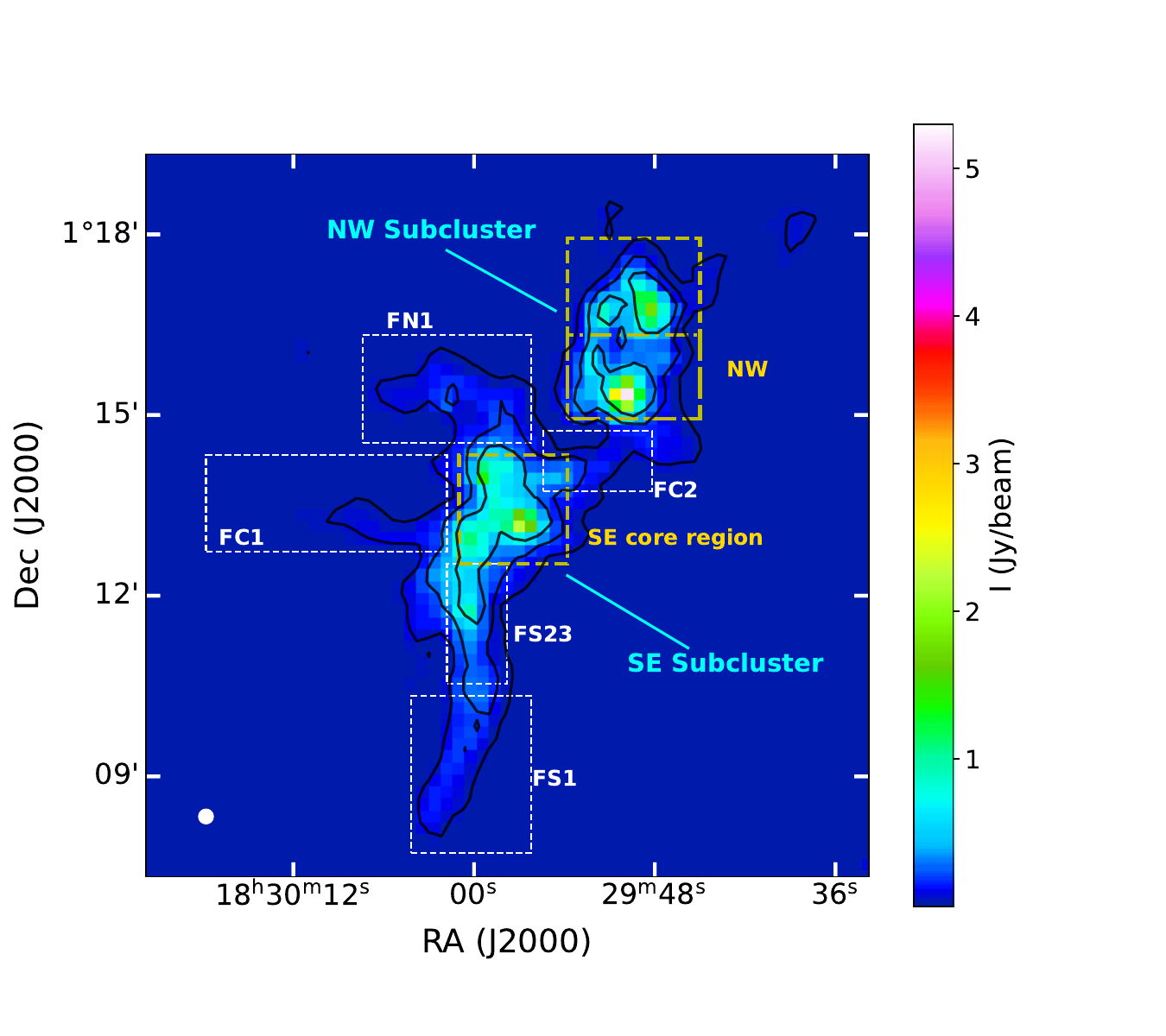}
    \caption{Map of Stokes-$I$ for Serpens Main. NW subcluster and SE subcluster,  along with the individual filaments (FS1, FS23, FC1, FC2, and FN1), are shown. The black contours represent the total emission intensity levels at $ I=0.05$, $ 0.2$, and $0.5$ Jy/beam. The filled circle at the bottom left corner shows the beam size (14\arcsec) of JCMT-850 $\mu$m. Emission intensity is highest toward the NW and SE core regions.}
    \label{fig:stokesI}
\end{figure*}

We bin the data by averaging the original 4\arcsec\, pixels over 3×3 pixel blocks to improve the independence of the data points. The resulting pixel scale is 12\arcsec, which is close to the JCMT beam size.
Figure \ref{fig:stokesI} shows the intensity map of Stokes $I$, where the filaments and core regions are shown. The NW and SE core regions show significantly enhanced emission intensity compared to the other filamentary structures (FS1, FS23, FC1, FC2, and FN1).  

Using Stokes $Q$ and $U$, we derive the observed polarized intensity ($PI_{\rm obs}$) at each pixel as
\begin{equation}
PI_{\rm obs} = \sqrt{Q^2 + U^2}.
\label{eq:PIobs}
\end{equation}

As the uncertainties in $Q$ and $U$, denoted by $\delta Q$ and $\delta U$, respectively, cause a bias in the polarized intensity, the debiasing of $PI_{\rm obs}$ is calculated as
\begin{equation}
PI = \sqrt{Q^2 + U^2 - 0.5(\delta Q^2 + \delta U^2)}.
\label{eq:PI}
\end{equation}

The uncertainty in $PI$ is calculated as
\begin{equation}
\delta PI = \sqrt{\frac{Q^2 \delta Q^2 + U^2 \delta U^2}{Q^2 + U^2}}.
\label{eq:dPI}
\end{equation}

Further, the polarization degree and its uncertainty are given by
\begin{equation}
P(\%)  = 100 \times \frac{PI}{I}, 
\label{eq:P}
\end{equation}
\begin{equation}
\delta P (\%) = 100 \times \sqrt{\frac{\delta PI^2}{I^2} + \frac{\delta I^2 (Q^2 + U^2)}{I^4}}. 
\label{eq:dP}
\end{equation}

The polarization angle is calculated as
\begin{equation}
\theta = 0.5 \tan^{-1} \left( \frac{U}{Q} \right),
\label{eq:theta}
\end{equation}
with uncertainty
\begin{equation}
\delta \theta = 0.5 \times \frac{\sqrt{U^2 \delta Q^2 + Q^2 \delta U^2}}{Q^2 + U^2}.
\label{eq:dtheta}
\end{equation}

Thus, the polarization degree and polarization angle are calculated for each pixel. We select data with $I/\delta I > 30$ and $P/\delta P > 2$ for a good signal-to-noise ratio (SNR). Further analysis is conducted with this SNR condition until specified separately.

\section{Magnetic fields} \label{sec:B_field}
\begin{figure*}
    \centering
    \includegraphics[width=\textwidth]{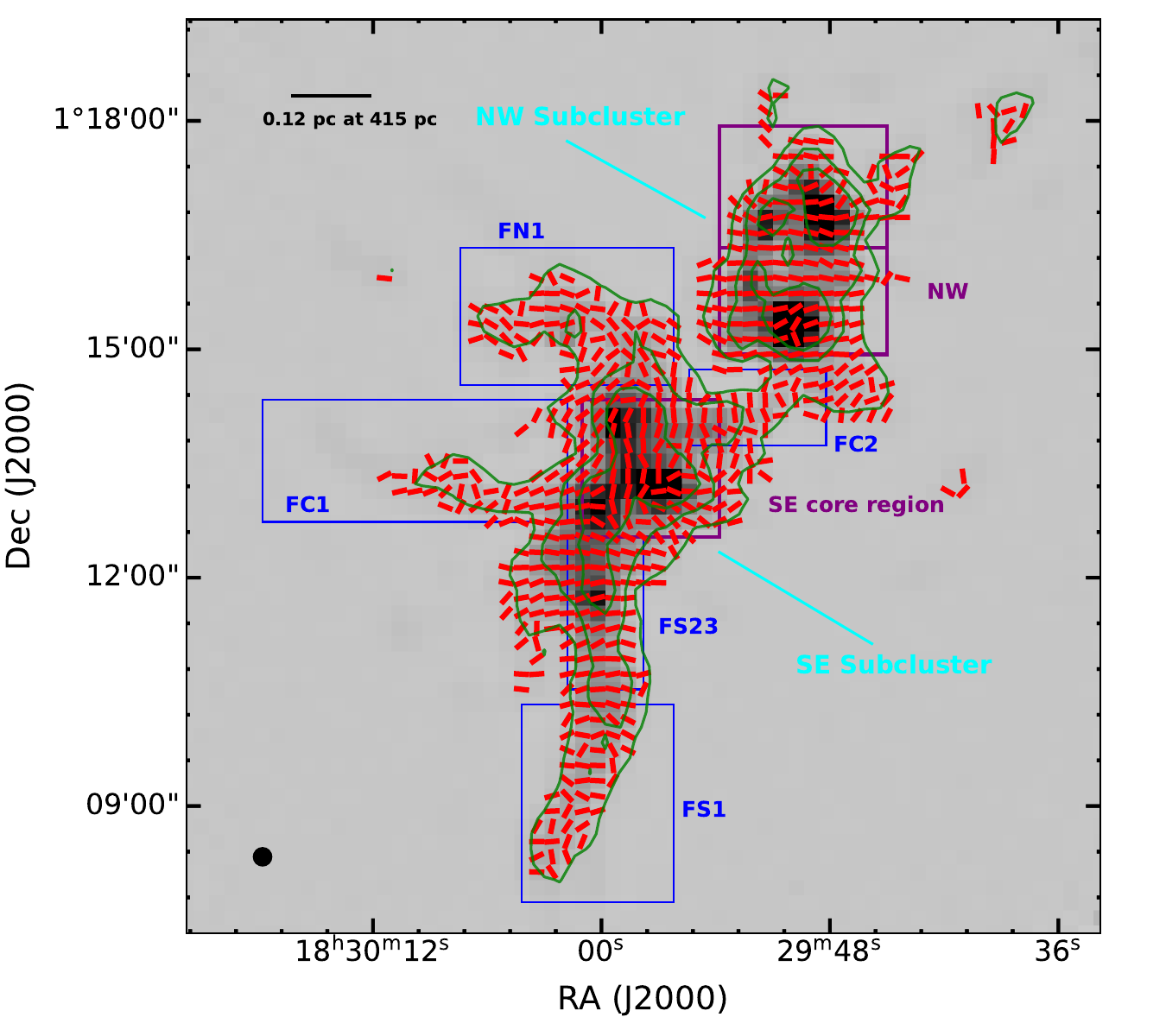}
    \caption{Morphology of $B_{\rm POS}$ overlaid on the total intensity map (Stokes $I$). All segments are plotted with equal length. Green contours represent the $I$ level at 0.05, 0.2, and 0.5 Jy/beam. The reference scale for 0.12 pc and the beam size are shown at the top left and bottom left of the plot. The $B$-field is oriented mainly in the east-west direction across most parts of the cloud.}
    \label{fig:B_orientation}
\end{figure*}

The plane-of-sky component of the magnetic field ($B_{\rm POS}$) is obtained by rotating the emission polarization angle by 90\degr. Figure \ref{fig:B_orientation} shows the $B_{\rm POS}$ morphology of Serpens Main. The same segment length is used for all segments to highlight the morphology of the $B$-field. A reference scale corresponding to 0.12 pc at the cloud distance of 415 pc is shown. 

The orientation of $B$-fields is predominantly in the east-west direction for the NW subcluster, while for the SE subcluster, various distinct orientations are observed. A complex $B$-field orientation is observed in the SE core region. \citet{2022ApJ...926..163K} conducted an in-depth investigation of the $B$-field orientations of filaments using the histogram of relative orientations (HRO) technique. The $B$-fields transition from being parallel to the filaments in less-dense filaments to being perpendicular to them in denser filaments; e.g., the orientation in filaments FS1, FS23, and FC2 is found to be perpendicular to the filaments, while for two filaments (FN1 and FC1), the orientation is parallel to the filament. Furthermore, applying the HRO technique to denser core regions revealed signatures of core formation, characterized by the reorientation of the $B$-field to a direction perpendicular to the density gradients.

\subsection{$B$-Field Strength Measurement Techniques} \label{B_strength}
The first method for estimating the $B$-field strength was introduced by \citet{1951PhRv...81..890D, 1953ApJ...118..116C}, which is known as the DCF technique. This method is based on the assumption that the $B$-field lines are distorted due to the propagation of incompressible transverse magnetohydrodynamics (MHD) waves (Alfv\'en waves) and that there is an energy equipartition between gas turbulence and magnetic turbulence. Assuming that the grains are aligned with the magnetic field, $B$-field fluctuations can be estimated from dust polarization angles. 
The strength of $B_{\rm POS}$ is then given by \citep[in cgs units;][]{2012ARA&A..50...29C}
\begin{equation}
B_{\rm POS} = \mathcal{Q} \times \sqrt{4 \pi \rho} \frac{\sigma_{\rm NT}}{\sigma_{\rm \theta}},
\label{eq:b_dcf}
\end{equation}
where $\rho$, $\sigma_{\rm NT}$, and $\sigma_{\rm \theta}$ are the gas mass density, one-dimensional nonthermal velocity dispersion, and polarization angle dispersion, respectively. Here $\mathcal{Q}\sim 0.5$ is a typical correction factor obtained from numerical simulations \citep{2001ApJ...546..980O}.
Equation \ref{eq:b_dcf} can be rewritten as \citep{2004Ap&SS.292..225C}   
\begin{equation}
B_{\rm POS} = 9.3 \sqrt{n({\rm H}_{2})}  \frac{\Delta V}{\sigma_{\rm \theta}}  ~(\mu {\rm G}),
\label{eq:b_dcf1}
\end{equation}
where $n({\rm H}_{2})$ is the number density in cm$^{-3}$, $\Delta V$ is the FWHM of the nonthermal velocity component in km/s, and $\sigma_{\rm \theta}$ is in degrees.

The original DCF method is known to exhibit uncertainties due to various effects related to accurate measurements of the angle dispersion and turbulent velocity dispersion (see, e.g., \citealt{2008ApJ...679..537F}).

Very recently, \citet[][hereafter \citetalias{2021A&A...647A.186S}]{2021A&A...647A.186S} noted that the assumption of incompressible turbulence made in the DCF method might not be realistic because ISM turbulence is compressible. They proposed a new method to estimate the $B$-field strength by including both incompressible and compressible modes as 
\begin{equation}
B_{\rm POS} = \sqrt{2 \pi \rho} \frac{\sigma_{\rm NT}}{\sqrt{\sigma_{\rm \theta}}},
\label{eq:ST}
\end{equation}
which becomes \citep{2021A&A...647A.186S, 2024A&A...684A.212B},
\begin{equation}
B_{\rm POS} = 1.8 \sqrt{n({\rm H}_{2})} \frac{\Delta V}{\sqrt{\sigma_{\rm \theta}}} ~(\mu {\rm G}).  
\label{eq:ST2}
\end{equation}
To obtain a pixel-to-pixel map of $B$-field strength, we construct maps of $\sigma_{\rm \theta}$, $n({\rm H}_{2})$, and $\Delta V$.

\subsection{Polarization Angle Dispersion} \label{angle_dispersion}
To calculate the dispersion of polarization angles, we use the method suggested by \citet{2021ApJ...913...85H}, which is based on the un-sharp masking technique \citep{2017ApJ...846..122P}. 
For each pixel in the resampled grid, we create a box of $3 \times 3$ pixels centered on that pixel. The pixel-box is chosen as $3 \times 3$, as this pixel-box size corresponds to approximately two to three beam sizes, as done in \citet{2017ApJ...846..122P, 2023ApJ...953...66N}. We first calculate the mean polarization angle $\left[\overline{\theta} = 0.5 \tan^{-1}(\overline{U}/\overline{Q})\right]$ using the mean Stokes parameters ($\overline{Q}$ and $\overline{U}$) in the $3 \times 3$ pixel-box and assign it to the central pixel of the box. 
We consider the supplementary angle for differences greater than 90\degr, due to the 180\degr\, ambiguity of polarization angles.

The polarization angle dispersion is calculated as 
\begin{equation}
\sigma_{\rm \theta} =   \sqrt{ \sum_{i=1}^{N}(\theta_i - \overline{\theta})^2 / N},
\label{eq:sigma_th}
\end{equation}
where $\theta_{i}$ is the polarization angle of the $i^{th}$ pixel and $N$ is the total number of pixels in the box. 
The value of $\sigma_{\rm \theta}$ is assigned to the considered pixel. We then move the box, repeat the entire process, and calculate $\sigma_{\rm \theta}$ for each pixel. 

\begin{figure*}
    \centering
    \includegraphics[width=0.52\textwidth]{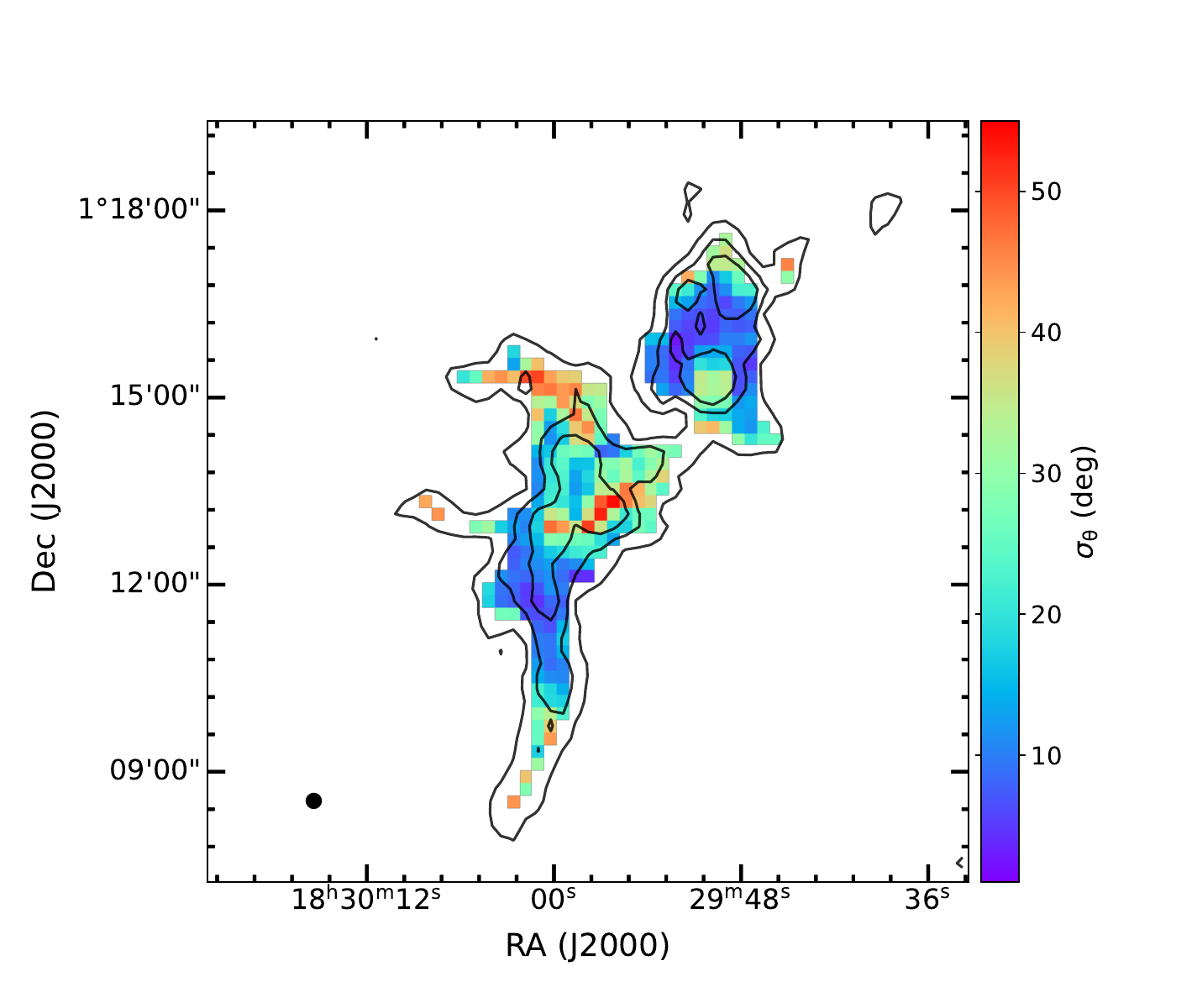}
	\includegraphics[width=0.45\textwidth]{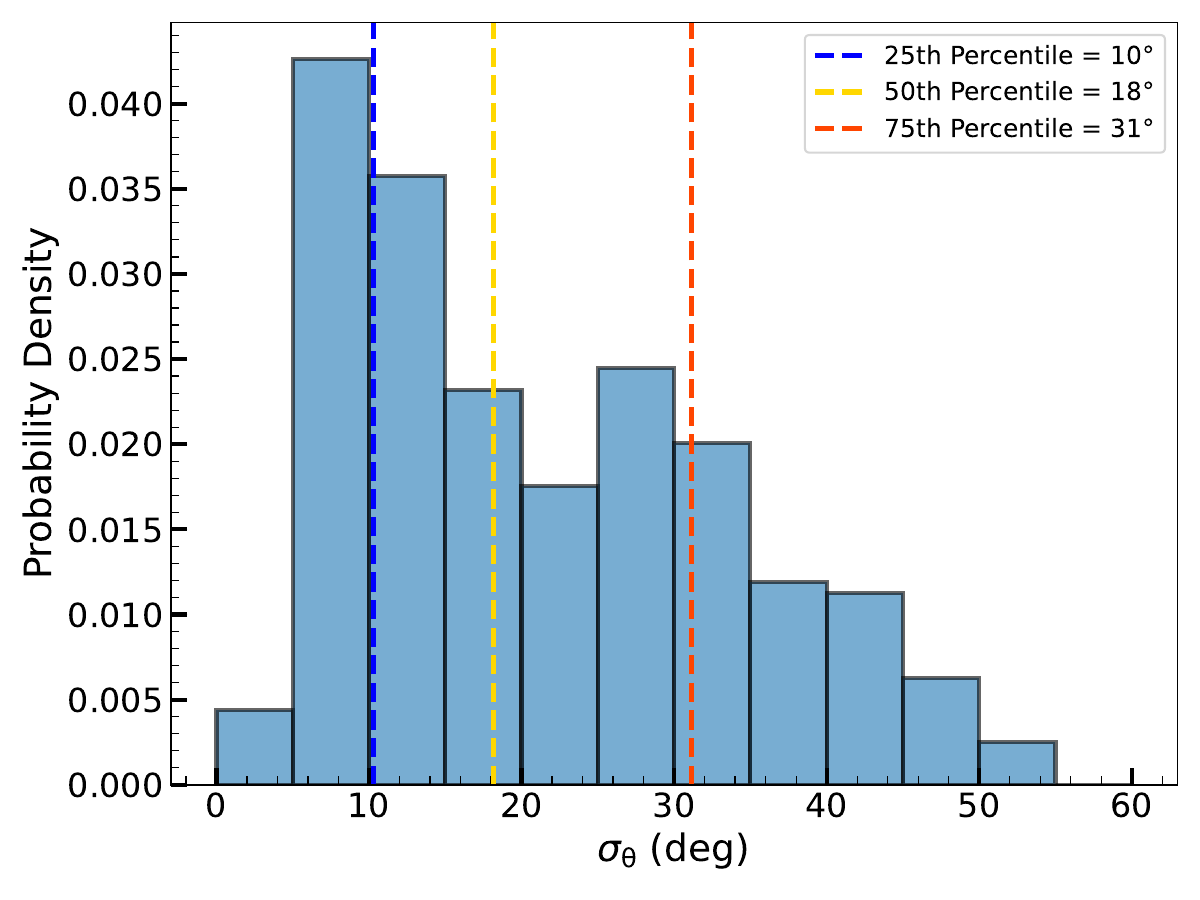}
    \caption{Left panel: map of the polarization angle dispersion. The contours correspond to $I$ levels of 0.05, 0.2, and 0.5 Jy/beam, and the black circle shows the beam size. Right panel: a histogram of polarization angle dispersion along with the 25$^{\rm th}$, 50$^{\rm th}$, and 75$^{\rm th}$ percentile values. The median value of $\sigma_{\rm \theta}$ is found to be 18\degr.}
    \label{fig:angle_dispersion}
\end{figure*}

The resulting map of $\sigma_{\rm \theta}$ and its distribution are shown in Figure \ref{fig:angle_dispersion}, left and right panels, respectively. The 25$^{\rm th}$, 50$^{\rm th}$, and 75$^{\rm th}$ percentile values of $\sigma_{\rm \theta}$ are marked with blue, yellow, and red vertical lines, respectively. The values of $\sigma_{\rm \theta}$ range from $\sim$ 3\degr\, to 54\degr\, with most values lying between $\sim$ 5\degr\, and 35\degr.  

\subsection{Column Density and Dust Temperature} \label{col_vol_density_dust_tem}
\begin{figure*}
    \centering
    \includegraphics[width=0.49\textwidth]{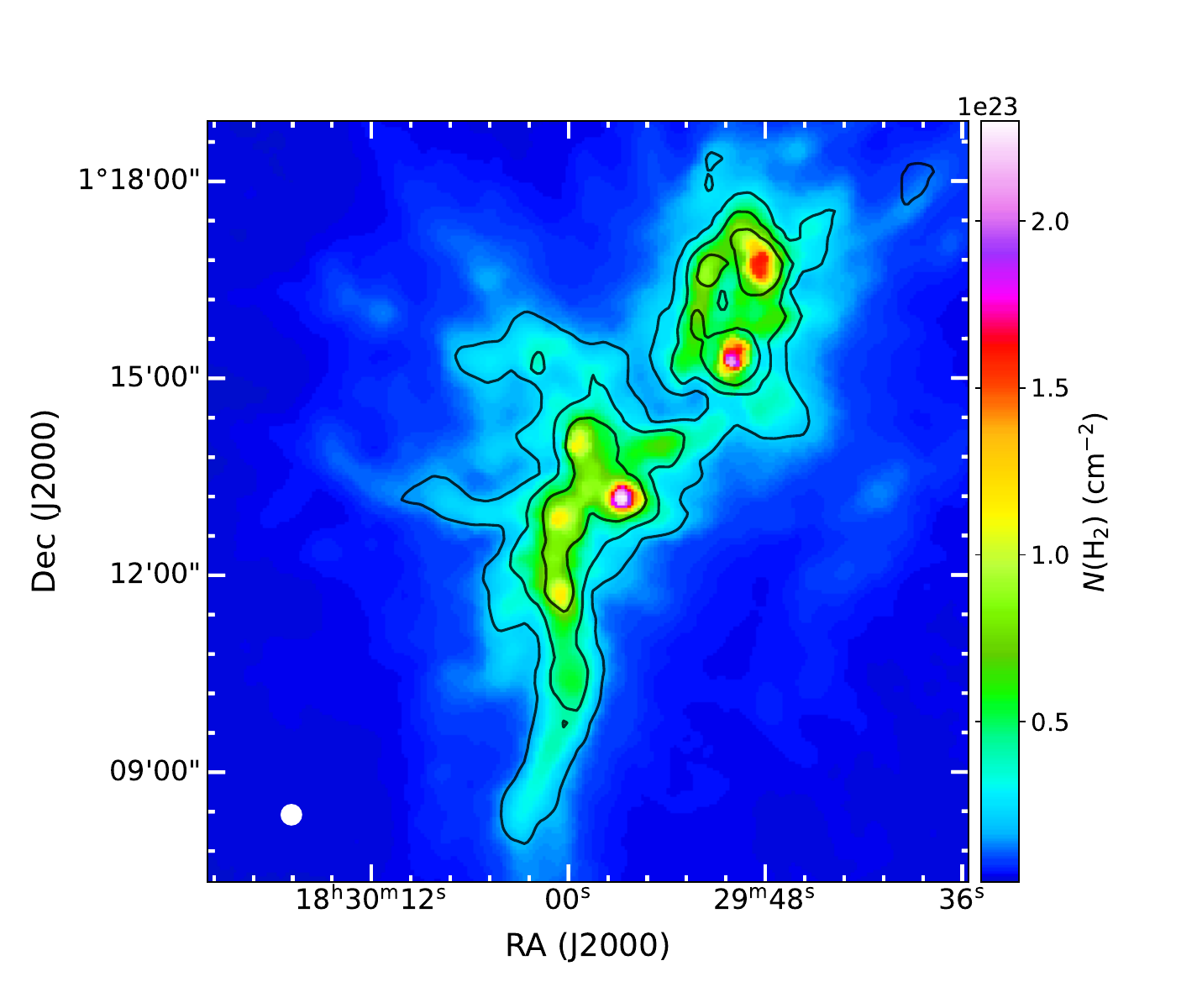}
    \includegraphics[width=0.49\textwidth]{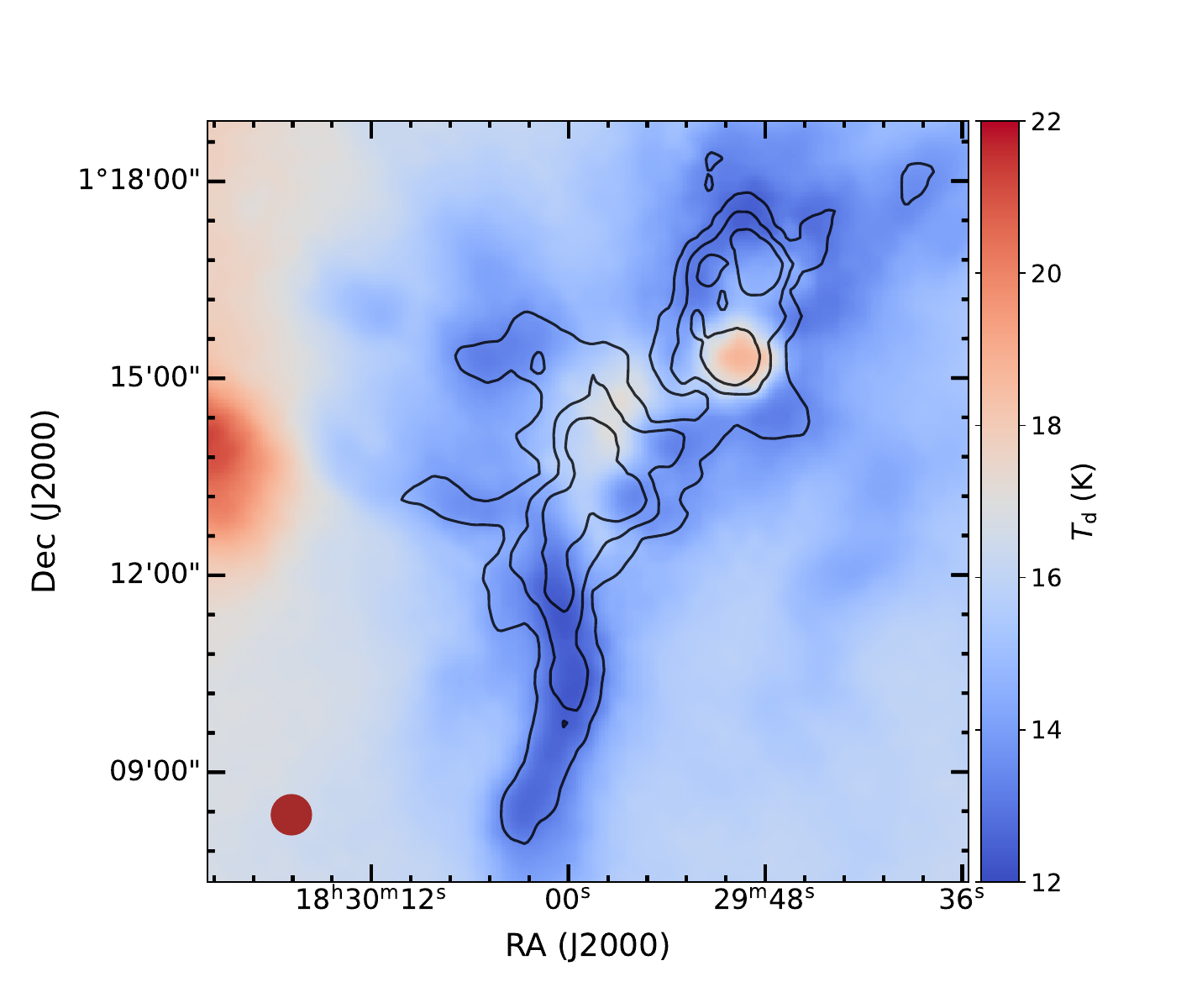}
    \caption{Map of column density (left panel) and dust temperature (right panel). The contours indicate $I$ levels of 0.05, 0.2, and 0.5 Jy/beam. A clear rise in dust temperature is seen in the densest NW subregion of the cloud.}
    \label{fig:column_density_and_dust_temp}
\end{figure*}

We take the map of column density [$N({\rm H}_{2})$] and dust temperature ($T_{\rm d}$) of Serpens Main from \citet{2021MNRAS.500.4257F} \citep[Herschel Gould Belt survey;][]{2010A&A...518L.102A}. The maps of $N({\rm H}_{2})$ and $T_{\rm d}$ are shown in Figure \ref{fig:column_density_and_dust_temp}. The column density map has a resolution of 18\arcsec.2, while the resolution for the dust temperature map is 36\arcsec.3 with a pixel size of $\sim$ 3\arcsec $\times$ 3\arcsec.   

The column density increases toward the core region of the cloud. Higher column densities are consistently observed around stronger emission intensity contours reflecting higher emission levels. The column density is highest in the SE core and NW subregions.  
Note that decreasing dust temperature toward denser regions shows that the attenuated interstellar radiation field is the primary source of dust grain heating and grain alignment, as shown in the FS1 and FS23 filaments of Serpens Main.
The core region of the NW subregion consists of the class 0 protostar Serpens SMM1 (RA: 18$^{\rm h}$29$^{m}$49$^{s}$.8, Dec: +01\degr15\arcmin20\arcsec.1), which is the most luminous source in the entire cloud of Serpens Main \citep{2014ApJ...797...76L}. The presence of an embedded source in the NW subregion results in enhanced dust temperatures within the dense core relative to its surroundings, as shown in Figure \ref{fig:column_density_and_dust_temp}. Among all subregions (FS1, FS23, FC1, FC2, FN1, NW, and SE core region), the highest temperature is found for the core region of the NW subregion, which is also associated with $I > 3$ Jy/beam.

\begin{figure*}
    \centering
    \includegraphics[width=0.49\textwidth]{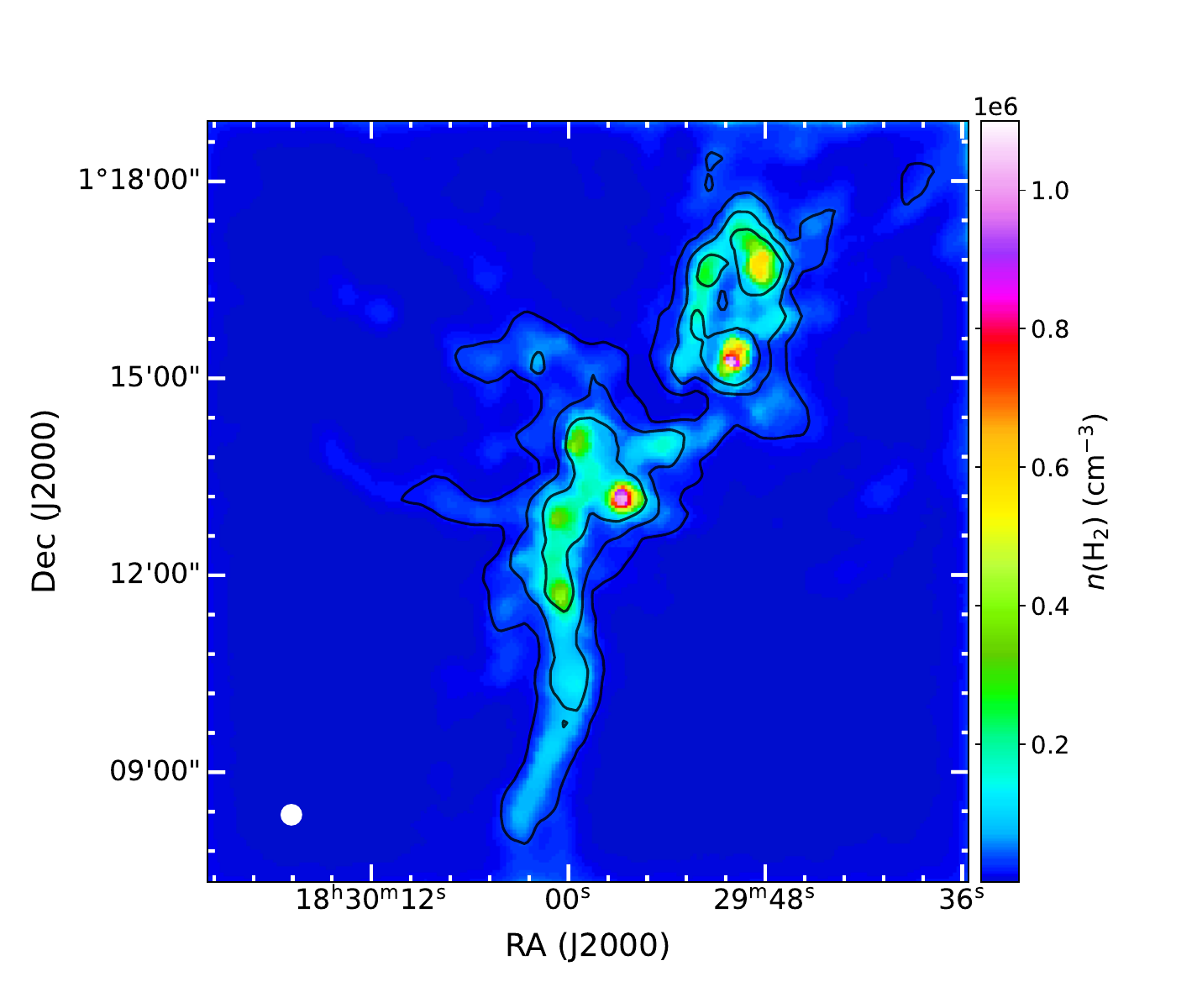}
    \includegraphics[width=0.49\textwidth]{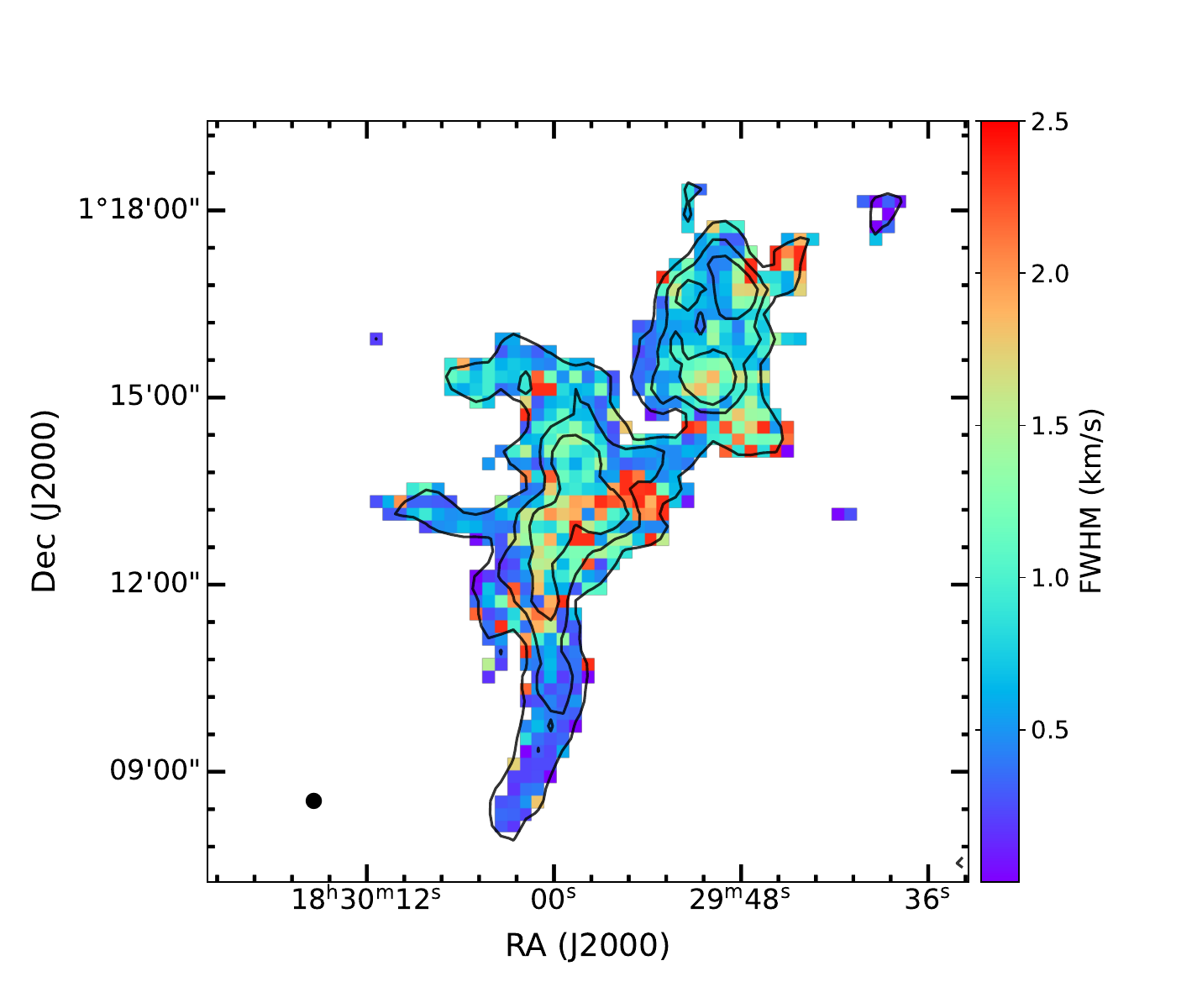}
    \caption{Left panel: map of volume density. Right panel: FWHM map of the nonthermal velocity component produced from $\rm N_{2}H^{+}$. The contours indicate $I$ levels of 0.05, 0.2, and 0.5 Jy/beam. The volume density increases toward the dense regions.}
    \label{fig:vol_density_vel_dispersion}
\end{figure*}

We derive the volume density from the column density map using the multiscale decomposition reconstruction (MDR) method introduced in \citet{2026ApJ...997..345Z}, which is based on the constrained diffusion algorithm \citep{2022ApJS..259...59L,2025arXiv250917369L}. In this approach, a 3D density field is reconstructed from a 2D column density map through structural analysis with the isotropic assumption. The MDR method has been tested with synthetic observations and applied by \citet{2026ApJ...997..345Z} to derive volume density maps. The volume density map derived from the column density of Serpens Main is shown in Figure \ref{fig:vol_density_vel_dispersion} (left).

\subsection{Nonthermal Velocity Dispersion} \label{vel_dispersion}

For gas velocity dispersion, we use the high-density gas tracer molecule $\rm N_{2}H^{+} (J = 1-0)$ data. The $\rm N_{2}H^{+} (J = 1-0)$ observations were made from the CARMA Large Area Star Formation Survey (CLASSy) by \citet{2014ApJ...797...76L}. $\rm N_{2}H^{+} (J = 1-0)$ consists of seven hyperfine components. To derive the velocity dispersion, we follow the method described by \citet{2014ApJ...794..165S} and simultaneously fit all seven hyperfine components, assuming the same dispersion and excitation conditions for all components. We model the full spectrum using the following equation: 
\begin{equation}
F(v) = \frac{2 k_{\rm B} v^2 \Omega(v)}{c^2} \times T_{\rm ex} (1-e^{-\tau(v)}) ,
    \label{eq:n2H_line}
\end{equation}
where $k_{\rm B}$ is the Boltzmann constant, $v$ is velocity, $\Omega$ is the synthesized beam area, $c$ is the speed of light in vacuum, and $T_{\rm ex}$ is the excitation temperature. Above, $\tau(v)$ is the optical depth, given by
\begin{equation}
\tau(v) = \tau_{0} \sum_{i=0}^{7} C_{\rm i} e^{-(v-V_{\rm lsr}-v_{\rm i})^{2}/2\sigma_{\rm v}^{2}}
    \label{eq:n2H_line_tau},
\end{equation}
where $\tau_{0}$ is the optical depth at the center of the line, $C_{\rm i}$ and $v_{\rm i}$ are the weighted strength and rest velocity of $i^{th}$ hyperfine component, $V_{\rm lsr}$ is the centroid velocity, and $\sigma_{\rm v}$ is the velocity dispersion. We adopt the value of $C_{\rm i}$ and frequency $\nu_{\rm i}$ from \citet{2009A&A...494..719P,2015PASP..127..266M} and the Cologne Database for Molecular Spectroscopy (CDMS) catalog available on Splatalogue\footnote{https://splatalogue.online/}. Here, $v_{\rm i}$ is calculated from $\nu_{\rm i}$ and the central frequency of the spectral line using the Doppler formula \citep{2014ApJ...797...76L}.

For each pixel, the velocity spectrum is fitted with the above model to obtain the velocity dispersion. Finally, the nonthermal velocity component is calculated as
\begin{equation}
\Delta V^{2} = \Delta V_{\rm total, N_{2}H^{+}}^{2} - 8\ln2\left( \frac{k_{\rm B} T_{\rm gas}}{m_{\rm N_{2}H^{+}}} \right), 
\label{eq:nonthermal_vel}
\end{equation}
where $\Delta V$ is the FWHM of the nonthermal component; $\Delta V_{\rm total, N_{2}H^{+}}$ is the FWHM of the $\rm N_{2}H^{+}$ spectral line; and the second term on the right-hand side is the thermal part, where $T_{\rm gas}$ is the gas temperature and $m_{\rm N_{2}H^{+}}$ is the mass of the $\rm N_{2}H^{+}$ molecule. We take the average temperature of $T_{\rm gas}= 15$ K, as \citet{2005ApJ...634..442S,2010A&A...523A..29D,2018ApJ...853..169D} mentioned the temperature range between 12-15 K, 10-20 K, and 17 K, respectively. However, the thermal contribution is found to be negligible. 
Using Equation \ref{eq:nonthermal_vel}, we calculate nonthermal velocity dispersion. The resulting map, shown in the right panel of Figure \ref{fig:vol_density_vel_dispersion}, is then reprojected, as described in Section \ref{sec:strength}, to match the spatial grid of our data.

\subsection{Map of $B_{\rm POS}$ Strength} \label{sec:strength}
We now use the maps of $\sigma_{\rm \theta}$, $n({\rm H}_{2})$, and $\Delta V$ obtained in Sections \ref{angle_dispersion}, \ref{col_vol_density_dust_tem}, and \ref{vel_dispersion} to derive the map of $B$-field strength. Since these maps have different pixel sizes or angular resolutions, we first smooth them to a common resolution \citep{2021ApJ...908...98G,2022ApJ...941...51H,2023ApJ...946....8T}. For this, we follow two steps:
\begin{itemize}
\item[i] Step 1: All maps are first reprojected to a common resolution. We use a common resolution consistent with the map corresponding to our dataset of $\sigma_{\rm \theta}$. 
\item[ii] Step 2: The reprojected maps are then smoothed by convolving with a Gaussian kernel: $\sigma_{kernel}= \sqrt{\sigma_{\rm target}^2 - \sigma_{\rm original}^2}$, where $\sigma_{\rm target}$ and $\sigma_{\rm original}$ are the $\sigma$ value of the target and original resolutions, respectively. The target resolution is set by the map with the lowest resolution among three maps \citep{2021ApJ...908...98G,2023ApJ...946....8T}. 
\end{itemize}

Using the resulting maps of $\sigma_{\rm \theta}$, $n({\rm H}_{2})$, and $\Delta V$, we then derive the map of $B_{\rm POS}$ using Equation (\ref{eq:ST2}). 
\begin{figure*}
\includegraphics[width=0.9\textwidth]{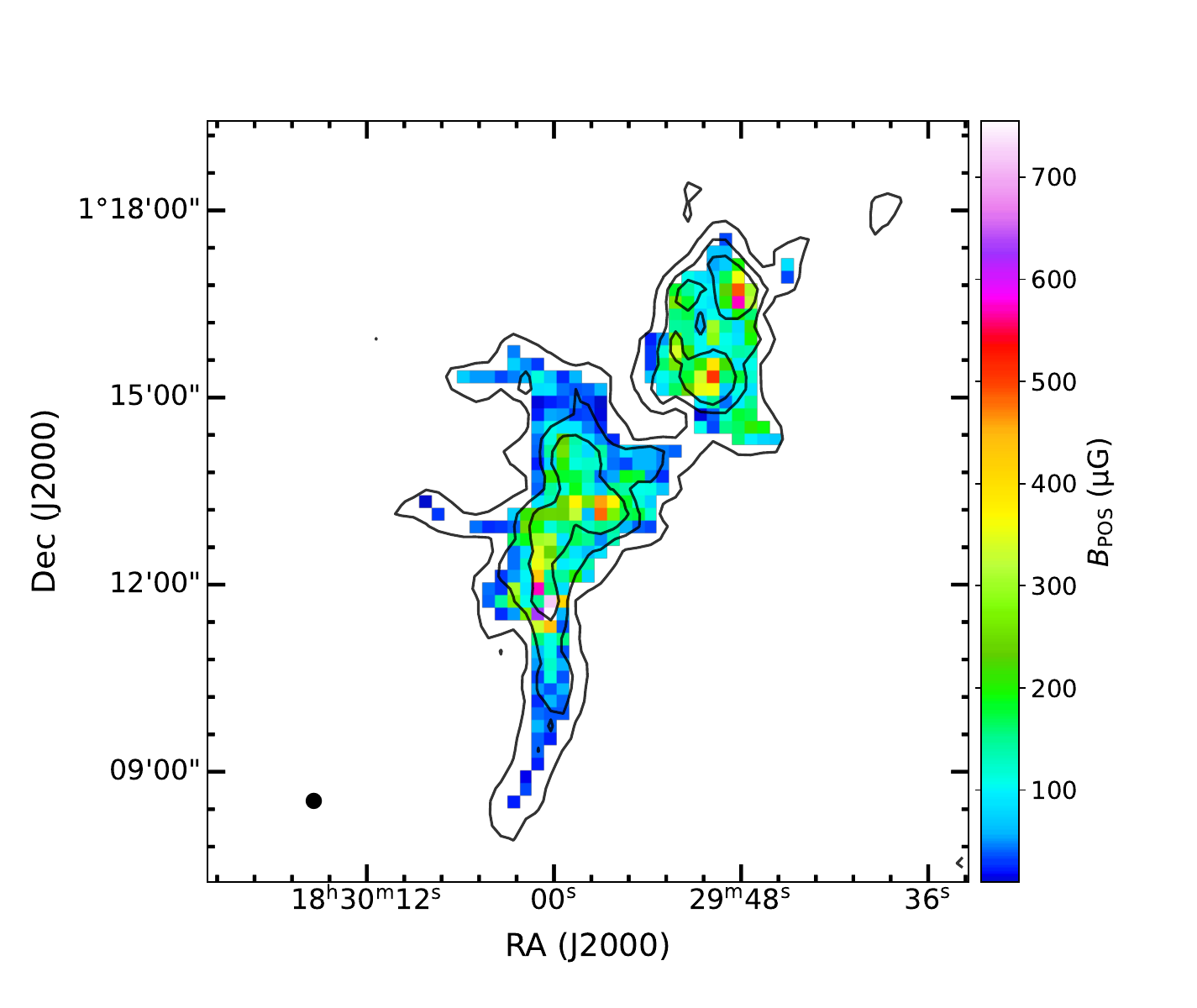}
    \caption{Spatial distribution of $B_{\rm POS}$ strength obtained from the \citetalias{2021A&A...647A.186S} method. The contours represent the $I$ level at 0.05, 0.2, and 0.5 Jy/beam. $B$-field strength tends to increase from the outer to the inner region, corresponding to an increase in  intensity contour levels.}
    \label{fig:B_strength_ST}
\end{figure*}
Figure \ref{fig:B_strength_ST} shows the map of $B$-field strength obtained from the \citetalias{2021A&A...647A.186S} method. The range of $B$-field strength extends up to $\sim$700 $\mu$G, increasing from the outer to the inner dense regions of the filament. The $B$-field strength is rather low ($\sim10-50 \ \mu$G) in the FS1 region and much stronger in the dense NW, SE core, and FS23 regions. 
The detailed dependence of $B$-field on the gas volume density is further discussed in Appendix \ref{apdx:B_n(H2)}. 
We also present the map of $B_{\rm POS}$ derived using the DCF method in Appendix \ref{apdx:B_dcf}; further analysis in this study is based on the map of $B_{\rm POS}$ obtained from the  \citetalias{2021A&A...647A.186S} method.

\subsection{Mass-to-flux Ratio}\label{lambda}
The relative role of the magnetic field to gravity in star formation is described by the ratio of the cloud mass ($M$) to the magnetic flux ($\Phi$). In terms of a critical value, the mass-to-flux ratio ($\lambda$) is defined as \citep{2004Ap&SS.292..225C}
\begin{equation}
\lambda = \frac{\left(M/\Phi\right)_{\rm observed}}{\left(M/\Phi\right)_{\rm critical}} = 7.6 \times 10^{-21}  \frac{N({\rm H}_{2})}{B_{\rm tot}} ,
\label{eq:mass_flux_ratio}
\end{equation}
where $N({\rm H}_{2})$ is in cm$^{-2}$, and $B_{\rm tot} = 1.3 \times B_{\rm POS}$ (in $\mu$G) \citep{2004Ap&SS.292..225C, 2021ApJ...907...88P, 2023ApJ...953...66N}. 

\begin{figure*}
    \centering
    \includegraphics[width=0.49\textwidth]{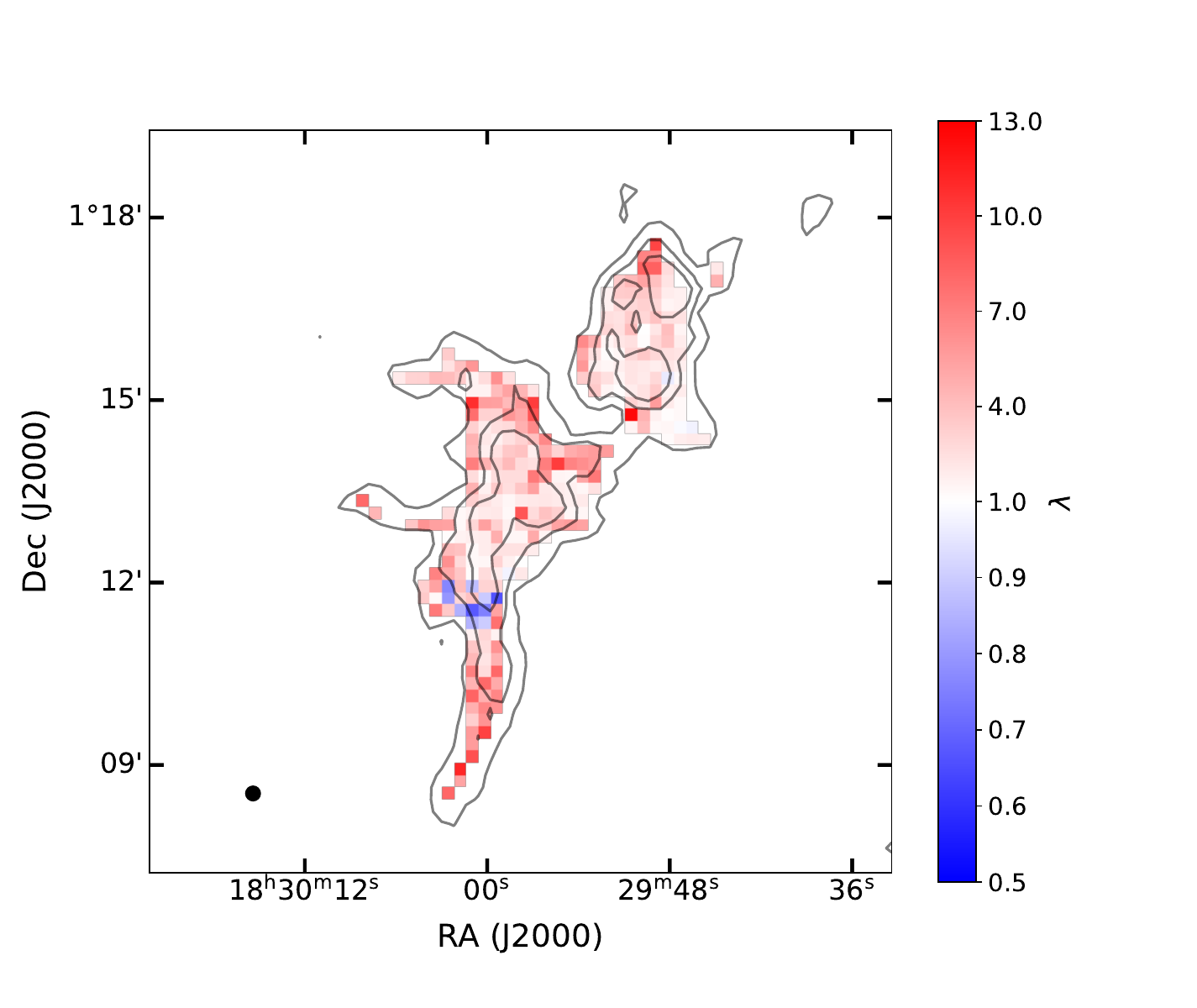}
    \includegraphics[width=0.49\textwidth]{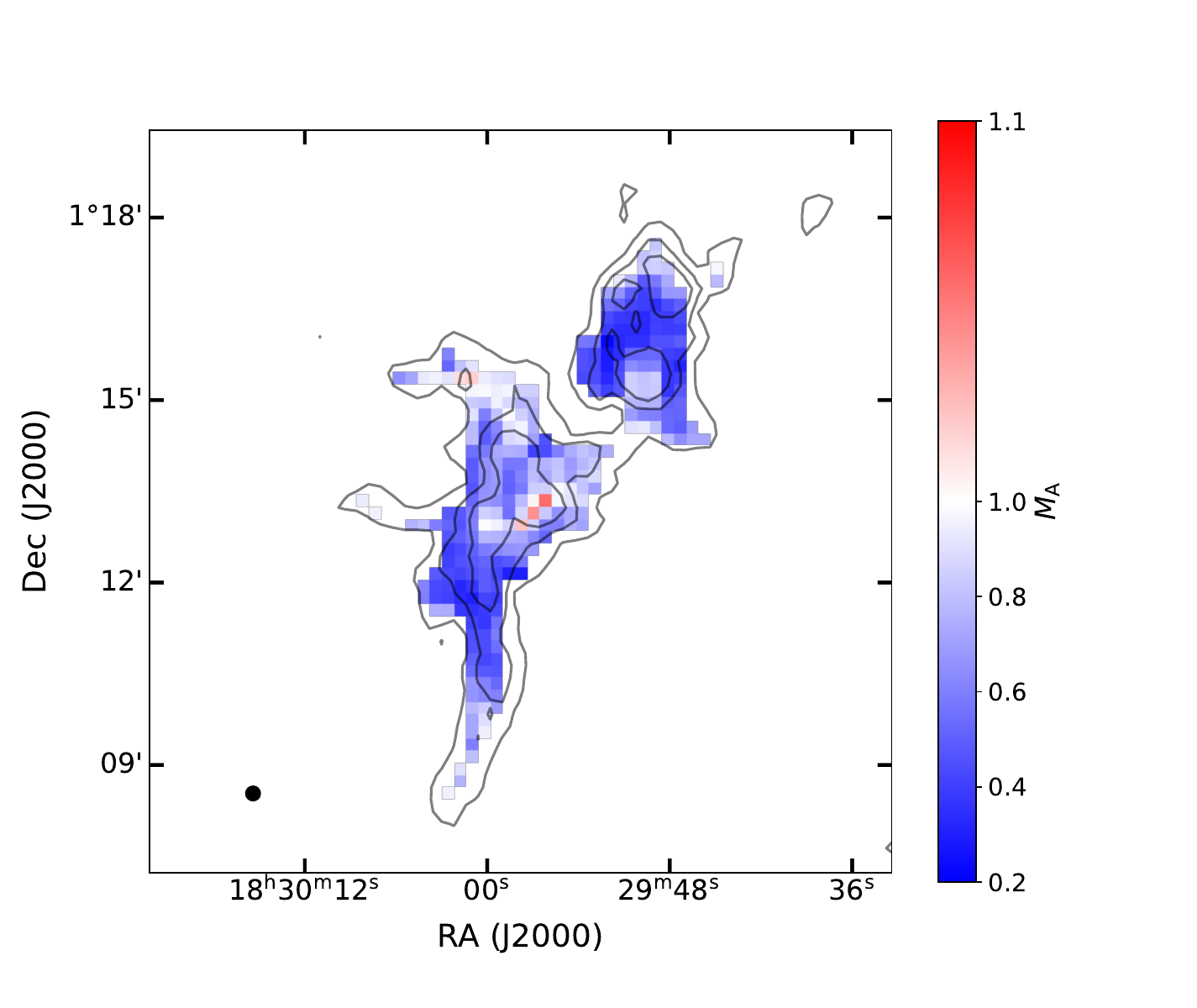}
    \caption{Maps of the mass-to-flux ratio (left panel) and Alfv\'en Mach number (right panel). The contours are at $I$ levels of 0.05, 0.2, and 0.5 Jy/beam. The cloud is predominantly supercritical and sub-Alfv\'enic.}
    \label{fig:mass_flux_ratio_and_alfven_mach}
\end{figure*}

The spatial distribution of the mass-to-flux ratio for Serpens Main is shown in Figure \ref{fig:mass_flux_ratio_and_alfven_mach} (left panel). The blue and red regions correspond to $\lambda < 1$ and $\lambda > 1$, respectively. The value of $\lambda < 1$ denotes a magnetically subcritical region, indicating that the magnetic field is strong enough to support the cloud against gravitational collapse. 
$\lambda > 1$, for all filamentary structures of the Serpens Main cloud, is indicative of magnetically supercritical regions. This suggests that the magnetic field in Serpens Main can no longer provide sufficient support against gravity.

\subsection{Alfv\'en Mach Number}  \label{alfven}
The Alfv\'en Mach number characterizes the importance of $B$-fields with respect to gas turbulence. The Alfv\'en Mach number of a gas is given by the ratio of the turbulent velocity to the Alfv\'en speed,
\begin{equation}
M_{\rm A} = \sigma_{\rm NT} / v_{\rm A}
\label{eq:alfven_mach1}
\end{equation}
where $\sigma_{\rm NT}$ is the nonthermal velocity dispersion of the gas and $v_{\rm A}$ is Alfv\'en velocity ($v_{\rm A} = B_{\rm tot} / \sqrt{4 \pi \rho}$) \citep{1953ApJ...118..116C}. Using Equation \ref{eq:b_dcf} and $B_{\rm tot}=1.3 \times B_{\rm POS}$, Alfv\'en Mach number becomes $M_{\rm A}=\sigma_{\rm \theta}/(1.3 \times Q)$, where $\sigma_{\rm \theta}$ is in radians. Here we are using $B_{\rm POS}$ obtained from the \citetalias{2021A&A...647A.186S} method; therefore, using Equation \ref{eq:ST2}, we have $M_{\rm A}=\sqrt{2 \times \sigma_{\rm \theta}}/1.3$, where $\sigma_{\rm \theta}$ is in radians. The map of Alfv\'en Mach number is shown in Figure \ref{fig:mass_flux_ratio_and_alfven_mach} (right panel). 
The regions with $M_{\rm A} <$ 1 and $M_{\rm A} >$ 1 represent sub-Alfv\'enic or super-Alfv\'enic gas, respectively. The sub-Alfv\'enic region shows that the magnetic pressure exceeds the turbulent pressure, while the turbulent pressure dominates in the super-Alfv\'enic case.
In most regions of the cloud, we find $M_{\rm A} <$ 1. This shows that turbulence is not dominant in Serpens Main, and hence the morphology of $B$-field is not influenced or affected by turbulent motion.

\section{Grain Alignment Mechanisms} \label{sec:dust_phy}
In this section, we study the properties of grain alignment derived from the observed polarization degree and polarization angles and interpret these properties using the modern RAT alignment theory, as outlined in BALLAD-POL I. We first analyze the correlations among key observational quantities, including the polarization degree ($P$), polarization angle dispersion function ($S$), total intensity ($I$), gas column density ($N({\rm H}_{2})$), and dust temperature ($T_{\rm d}$), to explore the role of varying local conditions on grain alignment and resulting dust polarization. We then calculate the key physical parameters of grain alignment using the RAT theory. Finally, we compare the resulting theoretical parameters with the observed quantities to test the grain alignment mechanisms.

\subsection{Polarization Degree with Total Intensity} \label{P-I} 
Previous studies have established that the dust polarization degree decreases with increasing emission intensity $I$, a.k.a. depolarization. Depolarization can be described by a power law, $P \propto I^{-\rm \alpha}$, in which $\alpha>0$ provides an important constraint on the variation of grain alignment efficiency and magnetic field properties in a cloud \citep{2015AJ....149...31J,2021ApJ...908..218H}. For example, in the case of $\alpha = 1$, the polarized intensity is independent of Stokes $I$, which indicates that grain alignment is only present in the outer region and becomes completely lost in the inner regions of the  cloud. However, the value of $\alpha = 0$ indicates that grain alignment and $B$-fields are uniform throughout the cloud. An intermediate value of $0 < \alpha < 1$ suggests that grain alignment is decreasing and/or the magnetic field becomes more turbulent toward the denser regions \citep{2021ApJ...908..218H}, where $\alpha=0.5$ corresponds to the case of completely random magnetic fields and uniform grain alignment (see, e.g., \citealt{2015AJ....149...31J}). 

In Serpens Main, \citet{2022ApJ...926..163K} derived $\alpha= 0.63$ by fitting a power law to the debiased polarization degree $P$, selecting data points that have $I/\delta I > 30$ and $P < 15\%$. Such an intermediate $\alpha$ implies that grain alignment is not completely lost in the inner regions of the filament.

However, the $\alpha$ value obtained from the conventional power-law fitting method tends to yield steeper slopes, which influences accurate understanding of physical origins underlying depolarization, such as grain alignment, in dense clouds \citep{2019ApJ...880...27P}. Therefore, to estimate $\alpha$, we use the Ricean mean method introduced in \citet{2019ApJ...880...27P}, which fits the variation of the nondebiased polarization fraction with $I$ by a Ricean mean model:
\begin{equation}
p'(I) = \sqrt{\frac{\pi}{2}} \sigma_{\rm p}(I) L_{\rm \frac{1}{2}}\left[-\frac{p(I)^{2}}{2\sigma_{\rm p}(I)^{2}} \right],
\label{eq:rice_mean}
\end{equation}
where $\sigma_{\rm p}(I) \approx \frac{\sigma}{I}$, and $L_{\frac{1}{2}}$ is the Laguerre polynomial of order $1/2$, and $p(I) = p_{\sigma} \left(\frac{I}{\sigma}\right)^{-\alpha}$ with $p_{\sigma}$ being the polarization fraction at the rms noise level of the data, $\sigma$. Here, $\sigma$ is the rms noise in Stokes $Q$ and $U$, which is calculated as 
\begin{equation}
\sigma = \frac{1}{2N} \sum_{i=1}^{N} \left(\sqrt{V_{\rm Q,i}} +   \sqrt{V_{\rm U,i}} \right),
\label{eq:sigmaqu}
\end{equation}
where $N$ is the total number of pixels in a central 3\arcmin\, diameter, and $V_{\rm Q,i}$ and $V_{\rm U,i}$ are the variances associated with each pixel. 

The above equation can be rewritten as
\begin{equation}
p'(I) = \sqrt{\frac{\pi}{2}} \left(\frac{I}{\sigma}\right)^{-1} L_{\rm \frac{1}{2}} \left[- \frac{p_{\rm \sigma}^{2}}{2} \left(\frac{I}{\sigma}\right)^{\rm 2(1-\alpha)}\right].
\label{eq:ricean_fit}
\end{equation}  

Our dataset is restricted to the central 3\arcmin\, diameter of best sensitivity around the center of both pointings \citep[NW and SE fields;][]{2022ApJ...926..163K}. We calculate $\sigma$ for both pointings  \citep{2021MNRAS.505..684P, 2021ApJ...918...85L, 2021ApJ...913...57K} and then take the mean as the final value for $\sigma$. The nondebiased polarization fraction is calculated using $p'=PI_{\rm obs}/I$ and further limited to $p' \leq 1$. Following the above method \citep{2019ApJ...880...27P, 2021ApJ...907...88P, 2021MNRAS.505..684P}, we derive the power-law index $\alpha$ and $p_{\rm \sigma}$. We did not notice a significant difference in the relation of $p'-I$ for both pointings; therefore, we combine the data of both pointings.  
Additionally, we apply a lower cut on Stokes $I$ to obtain a stable and statistically meaningful fit to Equation (\ref{eq:ricean_fit}). To ensure that the derived parameters ($\alpha$ and $p_{\rm \sigma}$) are not biased by a single empirical threshold, we repeat the fitting for several intensity cuts with $ 0.006 \leq I_{\rm cut} \leq 0.406$ Jy/beam, and derive the corresponding parameters and their associated uncertainties.
The derived value of $\alpha$ lies in the range of 0.62-0.78. The estimated weighted mean of $\alpha$ and $p_{\rm \sigma}$ are 0.68$\pm$0.01 and 0.53$\pm$0.02.

\begin{figure}
    \centering
    \includegraphics[width=\columnwidth]{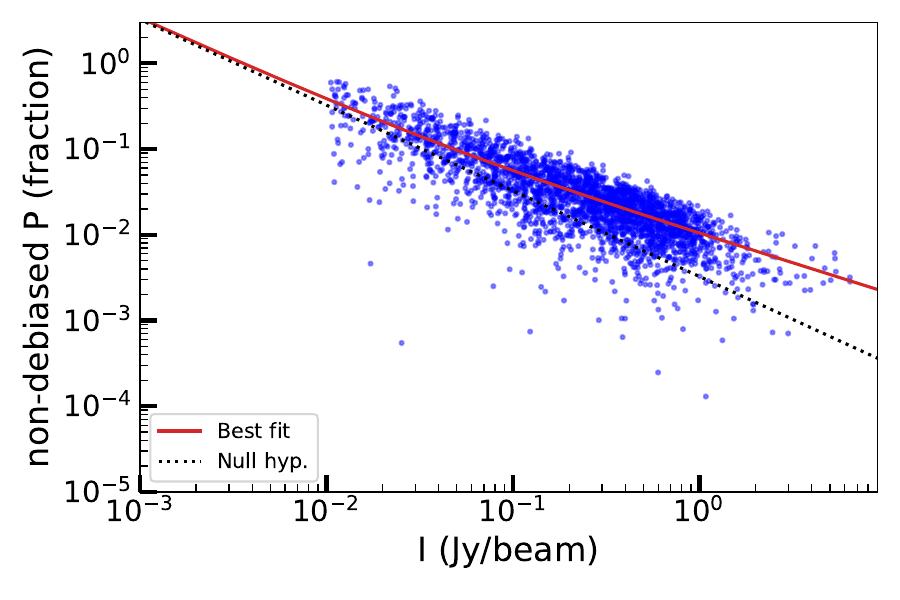} 
    \caption{Nondebiased polarization fraction ($p'$) as a function of total intensity for $I_{cut}$ = 0.0105 Jy/beam. The best-fit Ricean mean model (Equation \ref{eq:ricean_fit}) is shown by the red curve. The black dotted line indicates the null hypothesis, the expected behavior of nonaligned grains.}
    \label{fig:P-I}
\end{figure}

Figure \ref{fig:P-I} shows the variation of the nondebiased polarization fraction ($p'$) with the total intensity corresponding to $I_{cut}$ of 0.0105 Jy/beam. The best-fit model is shown by the red curve, and the black line shows the null hypothesis, $p'=\sqrt{\frac{\pi}{2}} \left(\frac{I}{\sigma}\right)^{\rm -1}$. This intermediate value of $\alpha=0.68$ implies that the grain alignment is not completely lost in the dense regions of Serpens Main. Interestingly, our derived slope using the Ricean mean method is slightly steeper than $\alpha=0.63$ obtained in \cite{2022ApJ...926..163K}, which implies that the power-law fitting using the debiased polarization is applicable for  Serpens Main. 

\subsection{Polarization Degree versus Column Density and Dust Temperature} \label{P_vs_N_T} 
To study how the polarization degree changes with local physical conditions, including gas density and radiation fields, we analyze the correlation of $P$ with $N({\rm H}_{2})$ and $T_{\rm d}$. 
\begin{figure*}
    \centering
    \includegraphics[width=0.49\textwidth]{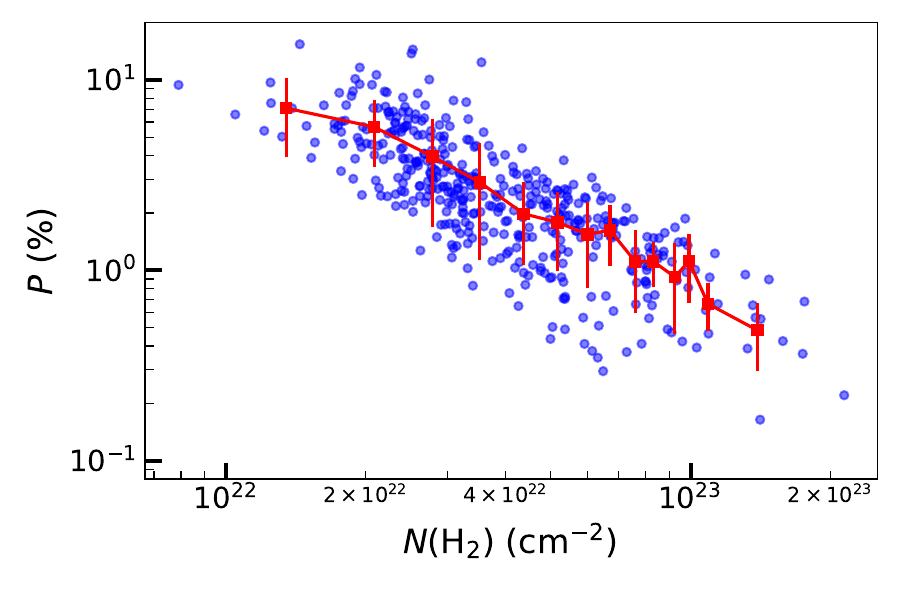}
    \includegraphics[width=0.49\textwidth]{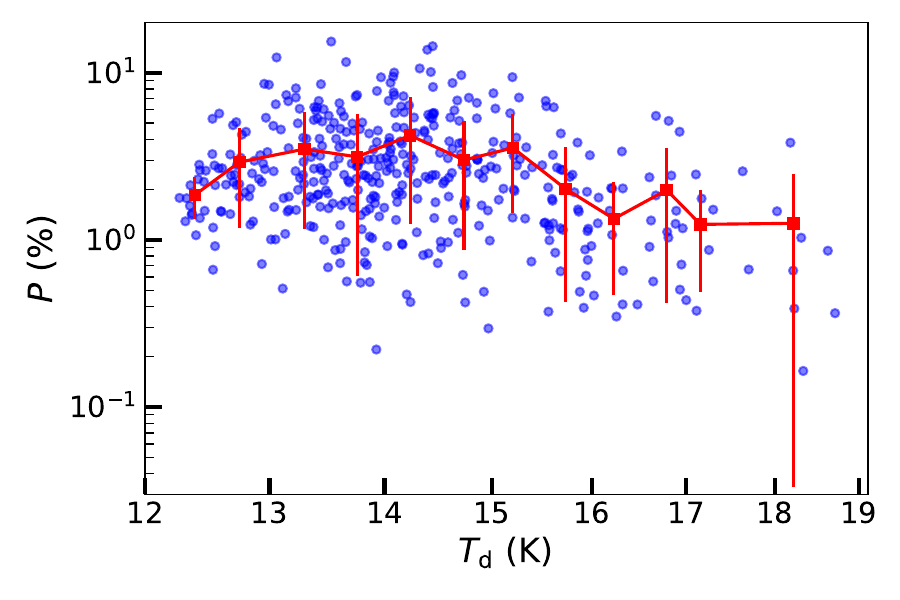}
    \caption{Polarization degree versus column density, $N({\rm H}_{2})$ (left panel) and dust temperature, $T_{\rm d}$ (right panel). The red curve shows the binned mean, with error bars denoting the standard deviation within each bin. $P$ decreases rapidly with increasing $N({\rm H}_{2})$, but it varies slowly with $T_{\rm d}$.}
    \label{fig:P_vs_column_dust}
\end{figure*}

Figure \ref{fig:P_vs_column_dust} (left panel) shows the variation of $P$ with the gas column density. The polarization degree decreases rapidly with increasing gas column density, a phenomenon commonly referred to as depolarization. This depolarization could arise from a gradual decrease in grain alignment efficiency toward the denser regions due to stronger gas randomization \citep{2021ApJ...908..218H}. 
Figure \ref{fig:P_vs_column_dust} (right panel) shows the variation of $P$ with the dust temperature. The mean polarization degree tends to slowly increase with temperature to $T_{\rm d}\sim 14$ K, then it decreases slightly for $T_{\rm d} > 15$ K. However, the decrease is not statistically significant. 

As the polarization degree depends on several parameters, including the grain properties, degree of grain alignment, and $B$-field tangling \citep{2024ApJ...965..183H}, we analyze the polarization angle dispersion function to disentangle the different effects.

\subsection{Magnetic Field Tangling: Polarization Angle Dispersion Function} \label{tangling}
To understand the importance of $B$-field tangling in depolarization, we next analyze the variation of the polarization angle dispersion function ($S$). Low values of $S$ are expected in regions with uniformly distributed polarization angles. However, if polarization angles are uncorrelated spatially, then a higher value of $S$ is expected \citep{2015A&A...576A.104P}. 

To derive $S$, we use the method described by \citet{2015A&A...576A.104P,2020A&A...641A..12P} for analyzing Planck polarization data. This method was later applied by  \citet{2021ApJ...908...98G,2023ApJ...953...66N,2024ApJ...974..118N} to {\it SOFIA} data. For each pixel at a location denoted by $\bx$, we define a circle at center $\bx$ with a radius of $\delta$. The radius $\delta$ is chosen to be nearly twice the beam size of JCMT/POL-2. Then for that pixel, we calculate the $S(\bx)$ as the rms of the angle differences between the pixel at $\bx$ and the pixel $i$ lying on the circumference of the circle, as given by
\begin{equation}
S^2(\bx) = \frac{1}{N} \sum_{i=1}^{N} [\theta(\bx) - \theta(\bx + \delta)]^2 ,
\label{eq:Sx}
\end{equation}
where $N$ is the total number of pixels lying on the circle. For $| \theta(\bx) - \theta(\bx + \delta)| > 90\degr$, we use the supplement of the angle.   

Earlier, we performed debiasing in the calculation of $P$ to account for the noise in the Stokes parameters $Q$ and $U$. Similarly, the bias on $S$ can be either positive or negative. Therefore, we calculate the variance of $S$ due to noise ($\sigma_{\rm S}$) as,       
\begin{equation}
\begin{split}
\sigma_{\rm S}^2(\bx) = \frac{\delta \theta^2(\bx)}{N^2 S^2}  \left(\sum_{i=1}^{N} [\theta(\bx + \delta) - \theta(\bx)]\right)^2 +  \\
\frac{1}{N^2 S^2} \sum_{i=1}^{N} \delta \theta^2(\bx + \delta)[\theta(\bx + \delta) - \theta(\bx)]^2
\end{split}
\label{eq:sigma_S}
\end{equation}
and use the debiasing as,
\begin{equation}
S_{\rm db}(\bx) = 
\begin{cases}
    \sqrt{S^2(\bx) - \sigma_{\rm S}^2(\bx)} \;\; & \text{if } S > \sigma_{\rm S} .\\
\end{cases}
\label{eq:Sdb}
\end{equation}
We refer to $S_{\rm db}$ as $S$ in the following analysis and derive a map of  polarization angle dispersion function.

\begin{figure*}
    \centering
    \includegraphics[width=0.49\textwidth]{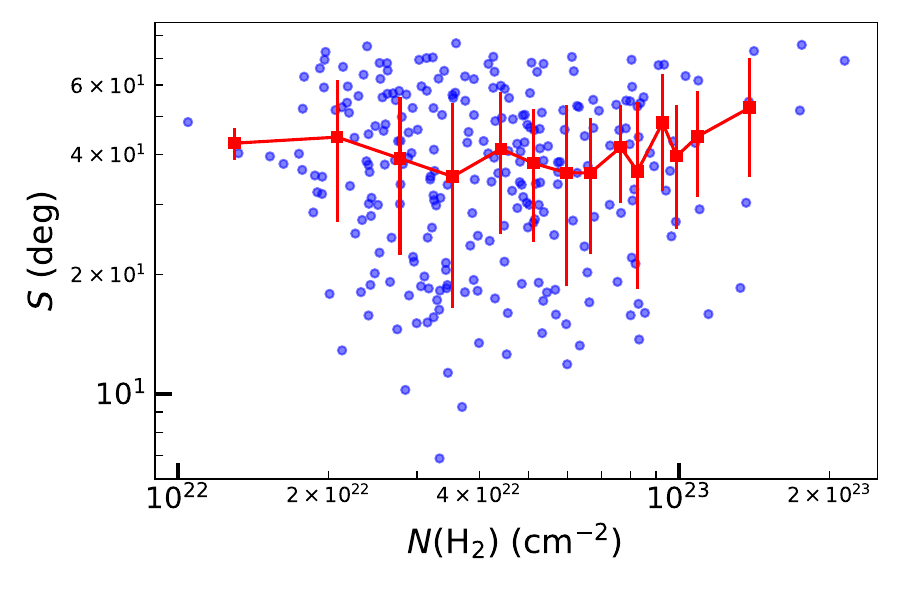}
    \includegraphics[width=0.49\textwidth]{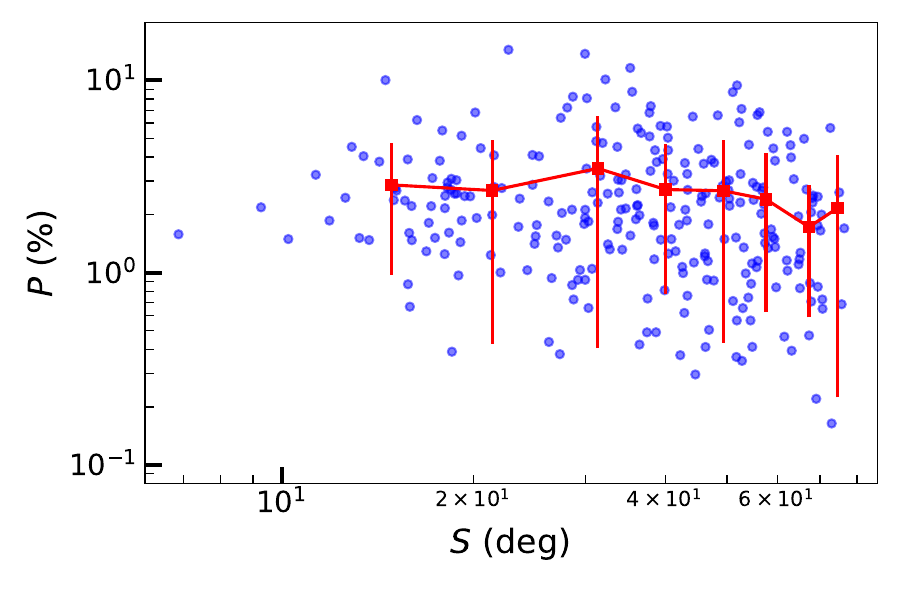} 
    \caption{Variation of $S$ with $N({\rm H}_{2})$ (left panel) and of $P$ with $S$ (right panel). No clear correlation is observed in either case.}
    \label{fig:S_vs_N_P_vs_S}
\end{figure*}

Variations of $S$ with $N({\rm H}_{2})$ and of $P$ with $S$ are shown in Figure \ref{fig:S_vs_N_P_vs_S}. The mean angle dispersion function $S$ does not exhibit a correlation with $N({\rm H}_{2})$ (left panel) and $P$ (right panel).  As a result, the $B$-field tangling is not a dominant cause for the depolarization observed in Figure \ref{fig:P-I}.    

\begin{figure*}
    \centering
    \includegraphics[width=0.49\textwidth]{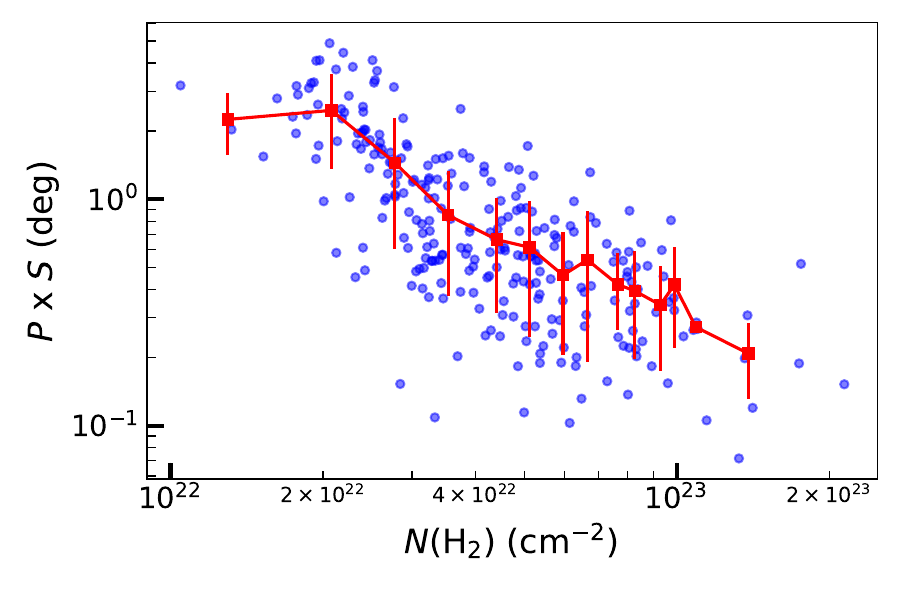}    
    \includegraphics[width=0.49\textwidth]{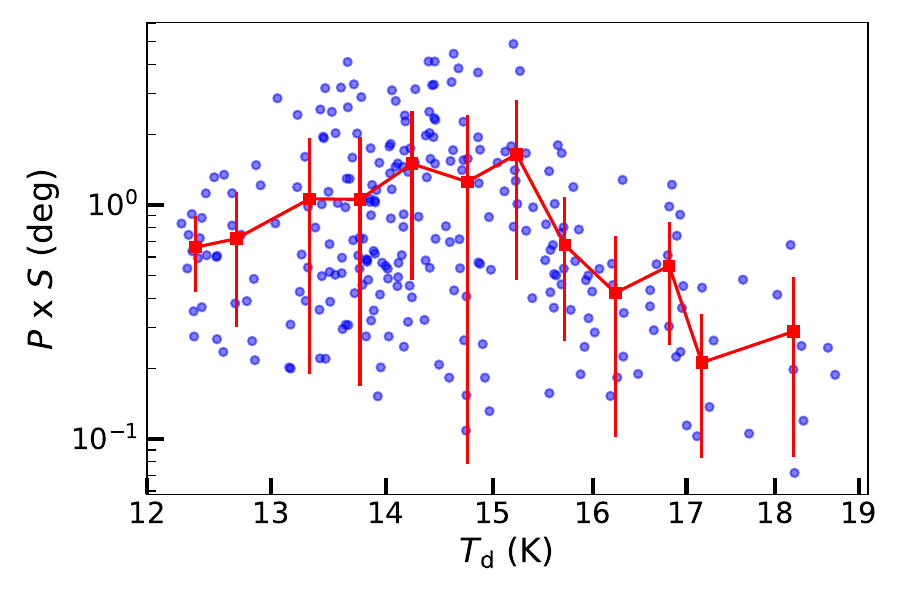} 
    \caption{Variation of net alignment efficiency, $P \times S$, with $N({\rm H}_{2})$ (left panel) and with $T_{\rm d}$ (right panel).}
    \label{fig:PxS_vs_other}
\end{figure*}

Figure \ref{fig:PxS_vs_other} shows the variation of $P \times S$ with $N({\rm H}_{2})$ (left panel) and $T_{\rm d}$ (right panel). $P \times S$ qualitatively characterizes the net alignment efficiency of grains observed along the line of sight \citep{2020A&A...641A..12P}. $P \times S$ tends to rapidly decrease with $N({\rm H}_{2})$ (left panel). Moreover, $P\times S$ first increases with $T_{\rm d}$ and then decreases for $T_{\rm d}>15$ K (right panel). The first trend is consistent with the RAT theory in which the grain alignment efficiency increases with $T_{\rm d}$, and the latter trend likely reflects the joint effect of higher gas density associated with higher $T_{\rm d}$ (see Figure \ref{fig:column_density_and_dust_temp}).

\subsection{Grain Alignment Size by RATs} \label{a_align}

To study the effect of grain alignment on  polarization properties, we quantify the key alignment parameters using RAT alignment theory.

For RAT alignment to be efficient,  grains must rotate suprathermally, i.e., their angular velocity spun up by RATs ($\omega_{\rm RAT}$) must be larger than  thermal angular velocity ($\omega_{\rm T}$) \citep{2008MNRAS.388..117H}. As described in \citet{2008MNRAS.388..117H, 2021ApJ...908..218H}, the minimum effective size of aligned grains (i.e., the alignment size--$a_{\rm align}$) is obtained using the suprathermal rotation threshold of $\omega_{\rm RAT} (a_{\rm align}) = 3 \omega_{\rm T}$. The alignment size is given by \citep{2021ApJ...908..218H,2022ApJ...928..102H} 
\begin{equation}
\begin{split}
a_{\rm align} \simeq 0.055 \, \hat{\rho}^{-1/7} \left(\frac{\gamma U}{0.1}\right)^{-2/7} \left(\frac{n_{\rm H}}{10^{3} \: cm^{-3}}\right)^{2/7} \\ \left(\frac{T_{\rm gas}}{10 \: K}\right)^{2/7} \left(\frac{\bar{\lambda}}{1.2 \: \mu m}\right)^{4/7} (1+F_{\rm IR})^{2/7} \; \mu m ,
\end{split}
\label{eq:a_align}
\end{equation}
where $\hat{\rho} = \rho_{\rm d} /(3$ $\rm g \: cm^{-3})$, with $\rho_{\rm d}$ as the dust mass density, $\gamma$ is the anisotropy of the radiation field, $\bar{\lambda}$ is the mean wavelength of the radiation field, and $F_{\rm IR}$ is the IR damping coefficient. In the above equation, $U=u_{\rm rad}/u_{\rm ISRF}$ denotes the radiation strength, where $u_{\rm rad}$ is the local radiation energy density and $u_{\rm ISRF}$ is the energy density of the local interstellar radiation field in the solar neighborhood. $n_{\rm H}=2 n({\rm H}_{2})$ is the number density of hydrogen atoms, and $T_{\rm gas}$ is the gas temperature, which can be approximated to the dust temperature for dense and cold environments. 

As shown in Equation \ref{eq:a_align}, the alignment size depends on the local radiation field ($\gamma$, $U$, $\bar{\lambda}$) and the local physical parameters of gas ($T_{\rm gas}$, $n_{\rm H}$). We assume $\gamma=0.3$, $\bar{\lambda}=1.2 \mu$m, similar to the mean wavelength of the interstellar radiation field, $T_{\rm gas} \approx T_{\rm d}$, and $U= \left(T_{\rm d}/16.4 \: \rm K \right)^{6}$ \citep{2021ApJ...908..159T}. In dense regions, IR emission damping is subdominant, and the factor of $(1+F_{\rm IR})^{2/7}\approx 1$ \citep{2022ApJ...928..102H}.
Using Equation (\ref{eq:a_align}), the alignment size is estimated from maps of the gas volume density and dust temperature. 

\begin{figure*}
    \centering 
    \includegraphics[width=0.49\textwidth]{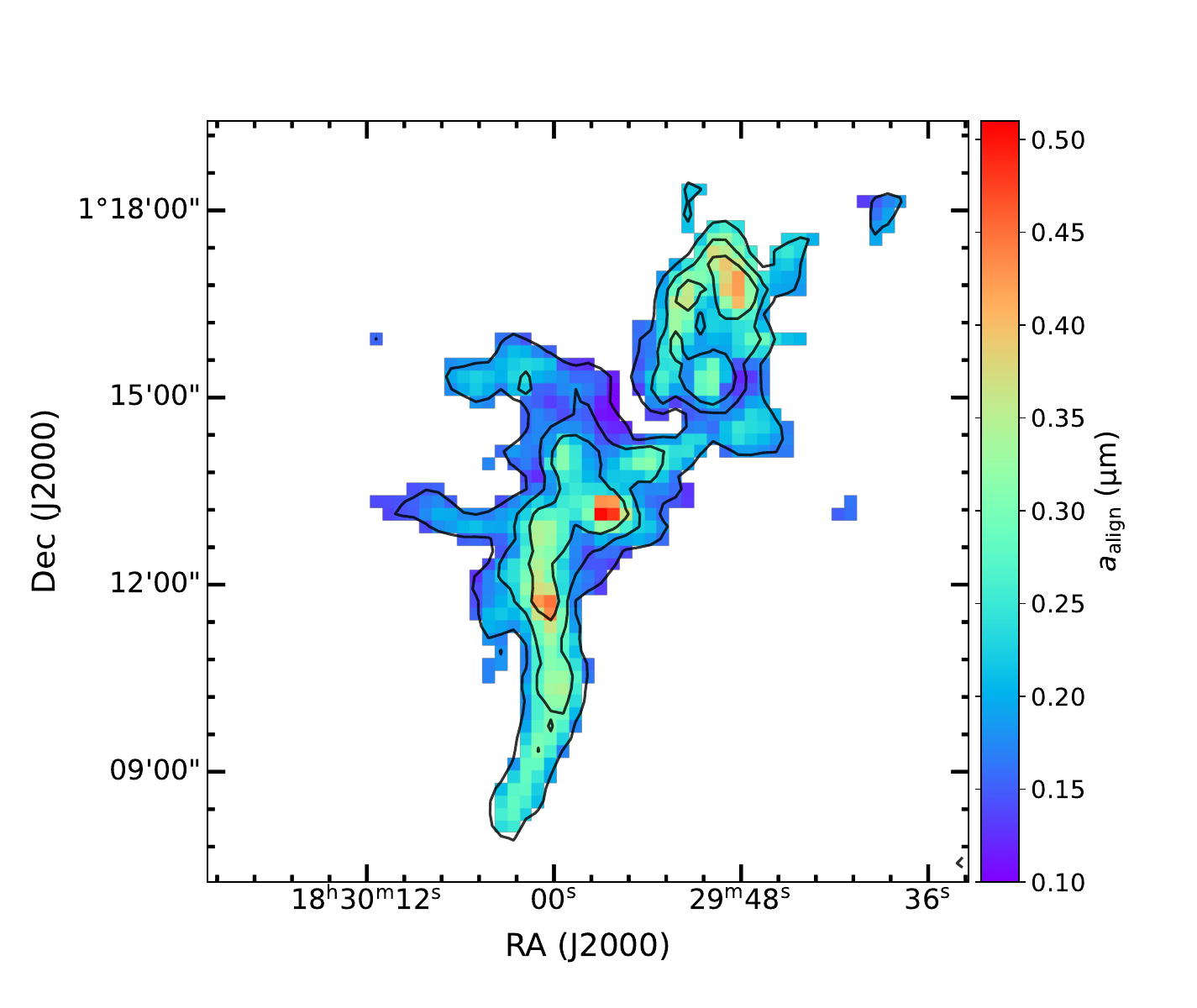} 
    \includegraphics[width=0.49\textwidth]{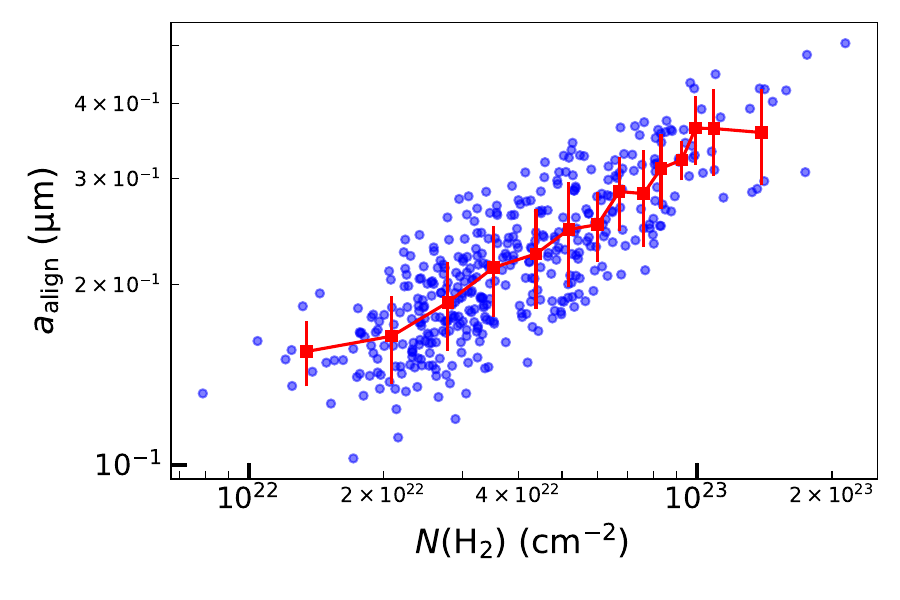}
    \caption{Left panel: map of  alignment size predicted by RATs, $a_{\rm align}$. The contours represent the $I$ levels at 0.05, 0.2, and 0.5 Jy/beam. Right panel: variation of $a_{\rm align}$ with $N({\rm H}_{2})$. The alignment size increases with $N({\rm H}_{2})$.}
    \label{fig:a_align}
\end{figure*}

\begin{figure*}
    \centering   
    \includegraphics[width=0.49\textwidth]{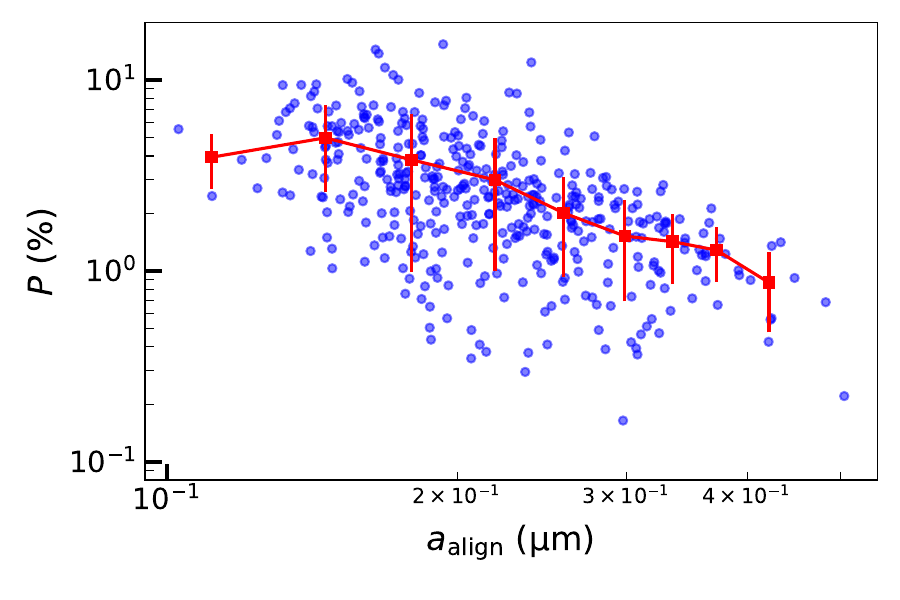}
    \includegraphics[width=0.49\textwidth]{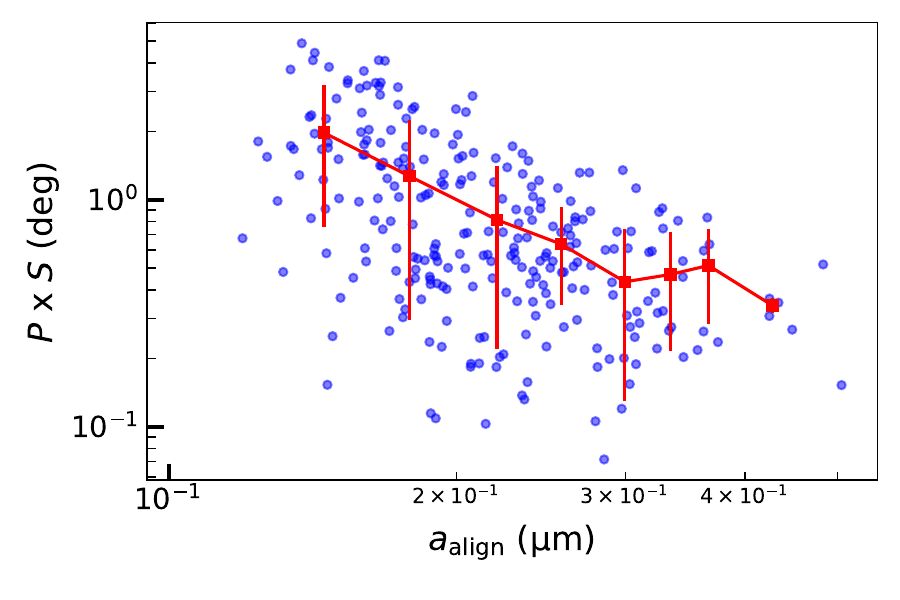} 
    \caption{Left panel: variation of  polarization degree ($P$) with alignment size, $a_{\rm align}$. Right panel: variation of  observed alignment efficiency, $P \times S$, with alignment size, $a_{\rm align}$. Both $P$ and $P \times S$ tend to decrease with an increase in $a_{\rm align}$.}
    \label{fig:a_align_vs_P_PxS}
\end{figure*}

The left panel of Figure \ref{fig:a_align} shows the map of $a_{\rm align}$. The alignment size $a_{\rm align}$ increases from the outer to the inner regions, reaching $\sim0.4-0.5 \ \mu$m in the densest regions. 
The right panel of Figure \ref{fig:a_align} shows the variation of $a_{\rm align}$ with $N({\rm H}_{2})$. The alignment size tends to increase with $N({\rm H}_{2})$. 
To see the variation of polarization degree and net alignment efficiency inferred from observations with $a_{\rm align}$, we show $P$ and $P \times S$ as functions of $a_{\rm align}$ in Figure \ref{fig:a_align_vs_P_PxS}. Both the polarization degree, $P$, and the  observed grain alignment efficiency, $P\times S$,  decrease rapidly with increasing $a_{\rm align}$. 
This indicates that the reduction in grain alignment efficiency predicted by RAT theory contributes to the observed depolarization. However, detailed modeling of dust polarization using the RAT theory including realistic magnetic fields is required to verify this interpretation.

\subsection{Effect of Magnetic Relaxation and  MRAT Mechanism} \label{delta_mag}
Magnetic inclusions embedded within dust grains can enhance the efficiency of RAT alignment \citep{2008ApJ...676L..25L, 2016ApJ...831..159H} due to the joint action of magnetic relaxation and RATs, which is termed the MRAT mechanism. A dimensionless parameter ($\delta_{\rm mag,sp}$) is introduced to describe the effect of magnetic relaxation over gas collisional disalignment \citep{2016ApJ...831..159H} and is given by  
\begin{equation}
\delta_{\rm mag,sp} = \frac{\tau_{\rm gas}}{\tau_{\rm mag,sp}} = 56 a_{-5}^{-1} \frac{N_{\rm cl} \phi_{\rm sp,-2} \hat{p}^{2} B_{3}^{2}}{\hat{\rho} n_{4} T_{\rm gas,1}^{1/2}} \frac{k_{\rm sp}(\Omega)}{T_{\rm d,1}} ,
\label{eq:delta_mag}
\end{equation} 
where $\tau_{\rm gas}$ and $\tau_{\rm mag,sp}$ are the gas collisional damping timescale and magnetic relaxation timescale, respectively, and $\delta_{\rm mag,sp} > 1$ indicates  effective magnetic relaxation. In equation (\ref{eq:delta_mag}), $a_{-5} = a / (10^{-5} \: \rm cm)$ with $a$ being the effective grain size, $N_{\rm cl}$ is the number of iron atoms per cluster, $\phi_{\rm sp,-2} = \phi_{\rm sp}/0.01$ with $\phi_{\rm sp}$ as the volume filling factor of iron clusters, $\hat{p} =p/5.5$ with $p\approx 5.5$ is the magnetic moment of an iron atom, $B_{3} = B_{\rm tot} / (10^{3} \: \mu \rm G)$, $n_{4} = n_{\rm H}/ (10^{4} \: \rm cm^{-3})$, $T_{\rm gas,1} = T_{\rm gas}/(10 \: \rm K)$, $T_{\rm d,1} =  T_{\rm d}/(10 \: \rm K)$, and $k_{\rm sp}(\Omega)$ is the function of grain rotation frequency ($\Omega$), which is of the order of unity \citep{2022ApJ...928..102H}. 

We use maps of the number density, magnetic field, and dust temperature, along with the typical values of $a = 10^{-5} \: \rm cm$, $N_{\rm cl} = 100$, and $\phi_{\rm sp} = 0.1$, and calculate $\delta_{\rm mag,sp}$ using Equation (\ref{eq:delta_mag}). We note that the choice of these parameters ($N_{\rm cl}$, $\phi_{\rm sp}$) is conservative, and it corresponds to about $30\%$ of the cosmic iron abundance in the form of embedded iron clusters \citep{2016ApJ...831..159H}. 

\begin{figure}
    \centering 
    \includegraphics[width=0.5\textwidth]{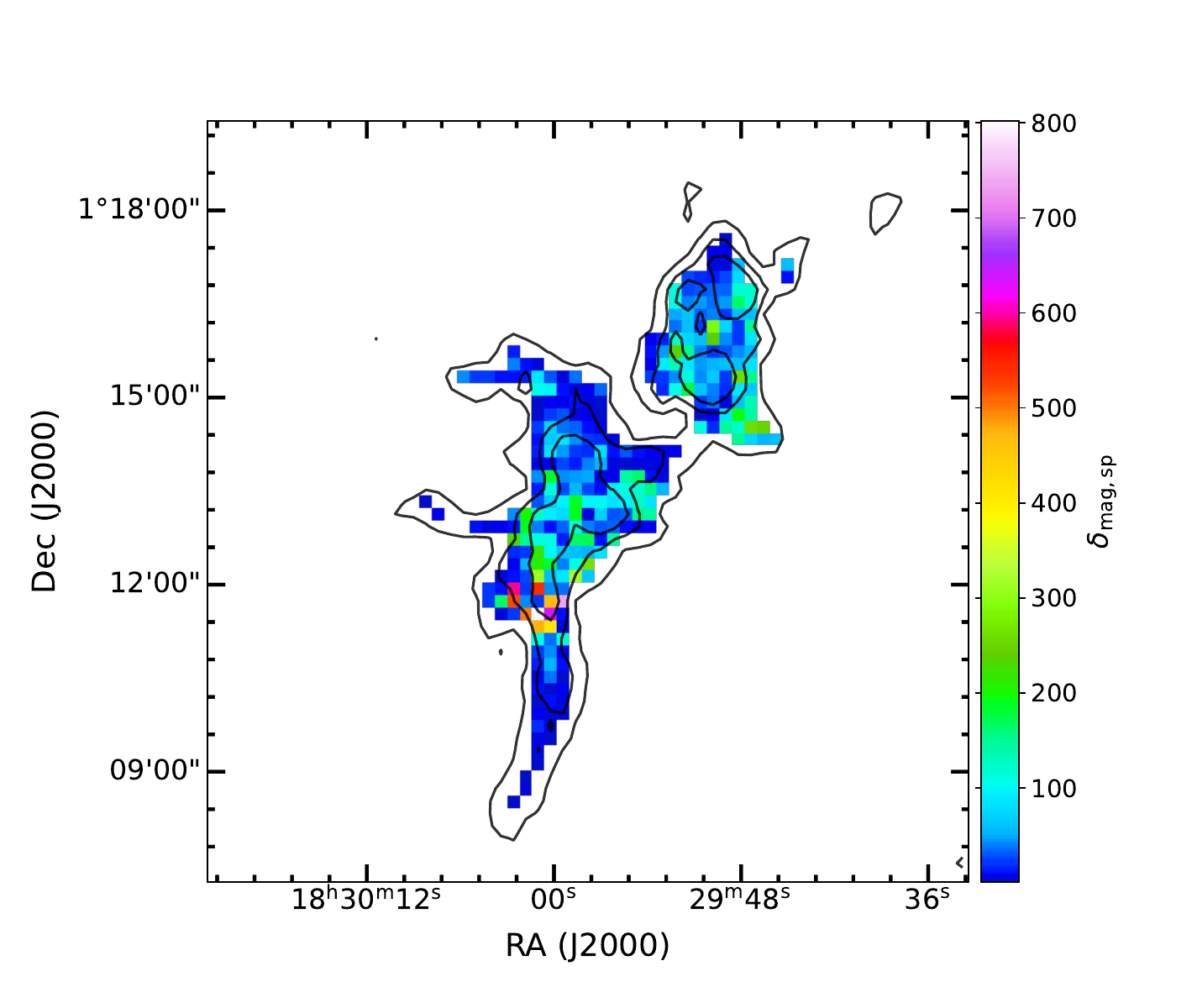} 
    \caption{The map of magnetic relaxation strength $\delta_{\rm mag,sp}$. The contours represent the $I$ levels at 0.05, 0.2, and 0.5 Jy/beam. $\delta_{\rm mag,sp} >1$ is found across most parts of the cloud.}
    \label{fig:delta_mag}
\end{figure}  

Figure \ref{fig:delta_mag} shows the map of $\delta_{\rm mag,sp}$. The value of $\delta_{\rm mag,sp}$ tends to be higher in regions with stronger magnetic fields (see Figure \ref{fig:B_strength_ST}), as expected from Equation (\ref{eq:delta_mag}). Some regions have $\delta_{\rm mag,sp} > 10$, indicating efficient magnetic relaxation, so that grains can be aligned efficiently via the MRAT mechanism \citep{2016ApJ...831..159H}. 

\begin{figure*}
    \centering
    \includegraphics[width=0.49\textwidth]{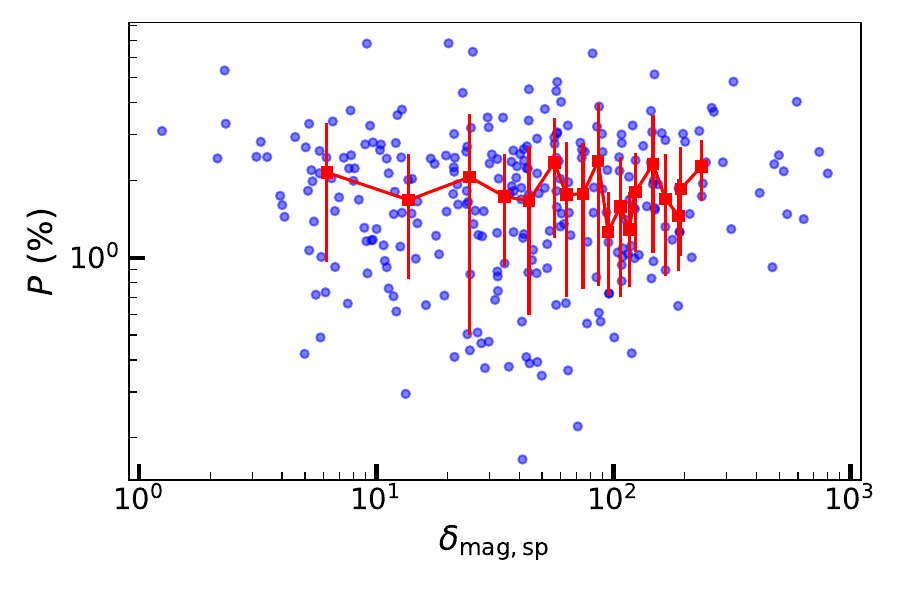} 
    \includegraphics[width=0.49\textwidth]{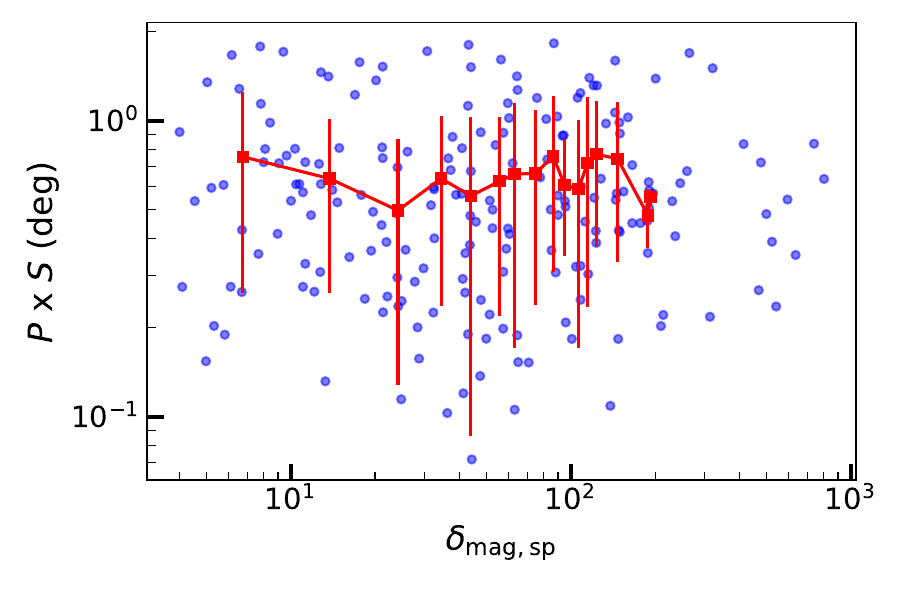}
    \caption{Variation of $P$ with  magnetic relaxation strength, $\delta_{\rm mag,sp}$ (left panel), and of $P \times S$ with $\delta_{\rm mag,sp}$ (right panel).}
    \label{fig:delta_mag_vs_P_PxS}
\end{figure*}

Figure \ref{fig:delta_mag_vs_P_PxS} shows the variation of $P$ and $P \times S$ as functions of $\delta_{\rm mag,sp}$, revealing no clear correlation. Such a lack of correlation does not imply that magnetic relaxation is ineffective. Rather, it may arise  from the competing effects between the alignment size, $a_{\rm align}$, and $\delta_{\rm mag,sp}$. Using the best-fit relation $B \propto n_{\rm H}^{0.6}$ from Figure \ref{fig:B_vs_n_A2}, we obtain $\delta_{\rm mag,sp} \propto B^{2}/n_{\rm H} \propto n_{\rm H}^{0.2}$. Thus, the increase in $\delta_{\rm mag,sp}$ is driven by the increase in $n_{\rm H}$. At the same time, $a_{\rm align}$ also increases with density, thereby counteracting the effect of enhanced magnetic relaxation on the resulting polarization.

\section{Discussion} \label{sec:diss}
\subsection{The Role of Magnetic fields}
Magnetic fields are believed to play a key role in the evolution of molecular clouds and the star formation process, in addition to gravity and turbulence \citep[see review by][]{2012ARA&A..50...29C}. In this paper, using dust polarization observations with JCMT/POL-2 toward Serpens Main, we first inferred $B$-field morphology and found that the $B$-field orientation is mostly along the east-west direction, or along the filamentary structure. However, in denser regions, the field morphology varies across the region, e.g., north-south orientation in some parts of the SE core region and more complex orientation in very dense clumps [$N({\rm H}_{2}) > 10^{23}$ cm$^{-2}$] of the SE core region. 

To quantify the relative role of $B$-fields to gravity and turbulence, we constructed a pixel-to-pixel map of $B_{\rm POS}$ strength. We found that the $B$-field strength obtained from the  \citetalias{2021A&A...647A.186S} method spans from $\sim 50 \ \mu$G in outer parts to $> 500 \ \mu$G in the densest parts of the cloud (see Figure \ref{fig:B_strength_ST}).
A stronger $B_{\rm POS}$ is observed within the region enclosed by the 0.5 Jy/beam intensity contour of $I$. An increase in the $B$-field strength with increasing density is also evident in the Serpens Main cloud, which is expected due to the freezing of magnetic field lines into the material (region 4 of \citealt{2022ApJ...926..163K}).
A power-law relation between the $B$-field strength and volume density yields $B_{\rm POS} \propto n({\rm H}_{2})^{0.60\pm0.10}$ (Figure \ref{fig:B_vs_n_A2}). \citet{1966MNRAS.133..265M} predicted a power-law exponent of 2/3 from theoretical consideration of a spherically symmetric contracting cloud. \citet{2010ApJ...725..466C} found a power-law exponent of 0.65 for $n({\rm H}_{2})> 300$ cm$^{-3}$ using total field strength including Zeeman observations. 
Our results for a power-law exponent of $0.60\pm0.10$ are consistent with \citet{1966MNRAS.133..265M,2019ApJ...871...98Z,2021ApJ...917...35M}.

The magnetic field usually varies from a few microgauss to a few milligauss \citep[high-mass SFRs;][]{2007MNRAS.382..699C} in the case of star-forming clouds.  Pixel-by-pixel maps of $B$-field strength have been derived for a few clouds in the past. For example, a range of $B$-field strength of 100-450 $\mu$G for 30 Doradus \citep[][at 154 $\mu$m]{2023ApJ...946....8T}, 0.2-2.5 $m$G for Rho Ophiuchus-A \citep[][at 154 $\mu$m]{2024A&A...690A.191L}, 100-2000 $\mu$G for Orion Molecular Cloud (OMC-1) \citep[][at 154 $\mu$m]{2021ApJ...908...98G}, 0.8-26.4 $m$G for OMC-1 \citep[][at 850 $\mu$m]{2021ApJ...913...85H}, 0.02-3.64 $m$G for Monoceros R2 \citep[][at 850 $\mu$m]{2022ApJ...941...51H}, 100-600 $\mu$G for G11.11-0.12 \citep[][at 214 $\mu$m]{2023ApJ...953...66N}, and 0.03-1 $m$G for G28.37+0.07 \citep[][at 214 $\mu$m]{2024ApJ...967..157L} were estimated. 
 
The relative importance of  magnetic field to gravity is quantified through the mass-to-flux ratio. We found that Serpens Main is mostly supercritical ($\lambda > 1$, see Figure \ref{fig:mass_flux_ratio_and_alfven_mach}, left panel), implying that most of the regions of Serpens Main seem to undergo a gravitational collapse. 
We evaluated the Alfv\'enic Mach number also and found that most regions of Serpens Main are sub-Alfv\'enic (see Figure \ref{fig:mass_flux_ratio_and_alfven_mach}, right panel), revealing the dominance of magnetic field over turbulence.

\subsection{Origins of Depolarization: $B$-field tangling versus grain alignment loss}
As in previous studies, we found evidence for depolarization in Serpens Main. The power-law fitting $P\propto I^{-\alpha}$ yields a slope of $\alpha = 0.68\pm0.01$ (see Section \ref{P-I}), indicating that grain alignment is still effective, but its efficiency is reduced. Below, we discuss the physical origins of depolarization  in detail.

In general, two possible explanations for the observed depolarization are the tangling of $B$-fields and a decrease in grain alignment efficiency toward denser regions. The effect of $B$-field tangling is discussed in Section \ref{tangling}, where we do not find any notable variation of $S$ with $N({\rm H}_{2})$. Thus, $B$-field tangling does not strongly affect the observed depolarization in Serpens Main.

To demonstrate the effect of grain alignment loss, we calculated the alignment size, $a_{\rm align}$, using RAT theory \citep{2021ApJ...908..218H} and found that $a_{\rm align}$ increases with increasing total intensity and gas column density. As predicted by numerical modeling \citep{2020ApJ...896...44L}, when the maximum grain size ($a_{\rm max}$) is fixed, an increase in the alignment size leads to a narrower size distribution of the aligned grains (i.e., a loss of grain alignment), thereby reducing the polarization degree. We note that the new effect of RATD due to internal stellar radiation, which might be another possible explanation for the depolarization in SFRs, is ineffective in Serpens Main; only the interstellar radiation field is responsible for RAT alignment \citep{2021ApJ...908..218H}.  Therefore, the dominant cause of the observed depolarization in Serpens Main is the decrease in grain alignment efficiency by RATs toward denser regions.

\subsection{Observational Results of Grain Alignment and Evidence for RAT Mechanisms} 
The negative correlation between $P$ and $N({\rm H}_{2})$ indicates  reduced grain alignment toward higher gas column densities. 
In RAT theory, dust grain alignment is driven by radiative torques produced from  anisotropic radiation. As predicted by RAT theory, the aligning radiation (i.e., the interstellar radiation field) becomes weaker toward the denser regions of a cloud; hence, the efficiency of RAT alignment decreases, causing a reduction in the polarization degree. 
However, in the presence of a bright embedded source in a cloud with stellar radiation dominating over interstellar radiation, the polarization degree is expected to increase due to  higher RAT alignment efficiency near the embedded source. For example, in a study of the power-law dependence of the polarization degree on the total intensity in the IC 348 star-forming region, \citet{2024ApJ...977...32C} found shallower power-law indices for cores with embedded protostars than for filamentary structures. 

In addition to the radiation field,  collisional randomization of grains in dense environments also affects the grain alignment efficiency. The competing effects of the radiative alignment and  collisional disalignment are incorporated through the alignment size.
We therefore derived the minimum size of aligned grains using RAT theory and analyzed the variation of observed parameters with $a_{\rm align}$.  
In general, an increase in $a_{\rm align}$ with $N({\rm H}_{2})$ is found. For a fixed maximum grain size, an increase in $a_{\rm align}$ directly leads to a decrease in  $P$ and $P \times S$ due to a narrower size distribution of aligned grains (\citealt{2020ApJ...896...44L,2021ApJ...906..115T}). This successfully explains the observed decrease in $P$ with $I$ and $N({\rm H}_{2})$. 

Here, we also note that $a_{\rm align} > 0.3 \ \mu$m in some regions of Serpens Main. This shows that the maximum grain size would be greater than $a_{\rm align}$, i.e., larger than the limit of maximum grain size in the diffuse ISM \citep{1977ApJ...217..425M}. This indicates the process of grain growth occurring in the cloud, which could be due to gas accretion and grain coagulation within the dense cloud.

According to RAT theory, a positive correlation is expected between the polarization degree and dust temperature, since the efficiency of grain alignment increases with the irradiation intensity (i.e., grain temperature; see \citealt{2020ApJ...896...44L}). 
We also observed a slight increase in $P$ with $T_{\rm d}$ up to $T_{\rm d}=14$ K. However, there is a slight decrease in $P$ for $T_{\rm d}>15$ K. We analyzed the variation of $S$ to infer the effect of $B$-field tangling. The polarization degree shows no significant variation with $S$, and the variation of $P \times S$ with $N({\rm H}_{2})$ and $T_{\rm d}$ follows a similar trend to that of $P$ with $N({\rm H}_{2})$ and $T_{\rm d}$. This indicates that tangled $B$-fields are not major contributors to the variation of $P-T_{\rm d}$.

\subsection{Implication for the MRAT Alignment Mechanism}
According to the MRAT mechanism, the alignment degree of dust grains depends on the grain properties (the size and magnetic susceptibility) and the local conditions (the gas, radiation field, and magnetic field).
Grains containing embedded iron inclusions (e.g., superparamagnetic grains) have enhanced magnetic relaxation from ordinary paramagnetic ones, so the joint effect of the enhanced magnetic relaxation and RATs increases the grain alignment degree by RATs only \citep{2016ApJ...831..159H}. 

\cite{2023ApJ...953...66N} first studied the MRAT effect using a map of $B$-field strength for G11.11-0.12 and found that magnetic relaxation is effective with $\delta_{\rm mag,sp}>1$ in this filament. \cite{2026arXiv260301899P} studied the MRAT effect by assuming the typical relation $B\propto n_{\rm H}^{2/3}$ and found the increase in the polarization degree with $\delta_{\rm mag,sp}$ in some regions of higher polarization with $\delta_{\rm mag,sp}>10$. In addition, $P$ decreases with an increasing $a_{\rm align}$ faster in the region with a low $\delta_{\rm mag,sp}$, but the decrease becomes much slower for $\delta_{\rm mag,sp}>10$ (see their Figure 13), suggesting evidence of magnetic relaxation that enhances polarization from grain alignment loss by RATs.

In this paper, we used the $B$-field map estimated from the \citetalias{2021A&A...647A.186S} technique to calculate $\delta_{\rm mag,sp}$, assuming 30\% of the iron abundance in the form of iron clusters in grains. We found an increase in $\delta_{\rm mag,sp}$ toward denser regions due to a stronger magnetic field. However, the polarization degree does not show a clear increase with $\delta_{\rm mag,sp}$. This likely arises from the competing effect of the grain alignment size, $a_{\rm align}$, because both  parameters increase toward regions of higher gas density. A detailed modeling of dust polarization using local properties and the estimated $\delta_{\rm mag,sp}$ using the DustPOL\_py code \citep{2021ApJ...906..115T} can help disentangle these effects. In particular, synthetic polarization observations based on the detailed MRAT theory are essential to disentangle the effects of dust physics and $B$-field tangling on  polarization properties. Using synthetic observations for protostellar cores, \cite{2024MNRAS.530..984G, 2025ApJ...980..105G} showed that the MRAT mechanism can efficiently align grains and reproduce high polarization degrees, $P\sim 20-30\%$, observed by ALMA.

\section{Conclusion} \label{sec:conclusion}
In this paper, we have studied the properties of the $B$-fields and grain alignment mechanisms for the Serpens Main molecular cloud using JCMT/POL2-850 $\mu$m polarization data. Our main results are summarized as follows:

\begin{enumerate}

\item The orientation of plane-of-sky $B$-fields is found mostly in the east-west direction, with more complex orientations in the dense regions of the cloud. We derive a map of $B_{\rm POS}$ using the \citetalias{2021A&A...647A.186S} method, with $B_{\rm POS}$ values reaching $\sim700 \mu{\rm G}$, and generally increasing from the outer regions toward the filament interior. The increase in the $B$-field strength with the gas volume density can be described by a power law with an exponent of $0.60\pm0.10$. 

\item From the derived $B$-field strength, we construct maps of the mass-to-flux ratio and Alfv\'en Mach number and found that Serpens Main is mostly supercritical with $\lambda > 1$ (i.e., gravitationally unstable region) and sub-Alfv\'enic with $M_{\rm A}<1$ (i.e., turbulence subdominant over magnetic fields). 

\item The polarization degree decreases with both the emission intensity and the gas column density ($N({\rm H}_{2})$), indicating well-known depolarization toward dense regions. A Ricean mean model fitting to the nondebiased polarization fraction vs $I$ yields a slope of $\alpha=0.68\pm0.01$, revealing the existence of grain alignment even in the denser regions. Based on various correlation analyses between $P$, $P\times S$, and gas column density, dust temperature, angle dispersion function, we found that the magnetic field tangling is not the primary cause of depolarization but reduced grain alignment is dominant.

\item We interpret the reduced alignment using the RAT alignment theory by calculating the minimum alignment size, $a_{\rm align}$, from the local physical conditions inferred from the observations. We found a correlation between $a_{\rm align}$ and $N({\rm H}_{2})$, and anticorrelations of $P$ and $P\times S$ with $a_{\rm align}$. This indicates that a narrower size distribution of aligned grains toward denser regions leads to a decrease in the polarization degree, in agreement with the predictions of RAT alignment. An indication of grain growth is seen in some regions of the cloud. 
 
\item Using the map of $B$-field strength from the \citetalias{2021A&A...647A.186S} method, we  calculate the magnetic relaxation strength $\delta_{\rm mag,sp}$. The high values of $\delta_{\rm mag,sp} > 10$ observed in the dense regions of the cloud indicate enhanced magnetic relaxation. However, no clear increase in the polarization degree is observed. This likely arises from the competing effect of an increase in the alignment size, $a_{\rm align}$, toward these regions.

\end{enumerate}

\begin{acknowledgments}
We are grateful to the anonymous reviewer for valuable comments and suggestions that helped improve the paper.
The initial work was carried out at the Physical Research Laboratory (PRL), supported by the Department of Space, Government of India.
The authors would like to thank Ms. Nguyen Chau Giang for valuable discussions regarding the density map. 
T.H. acknowledges the support from the major research project (No. 2026183200) from Korea Astronomy and Space Science Institute (KASI).
P.N.D. and N.B.N. are funded by the National Foundation for Science and Technology Development (NAFOSTED) under project code 103.99-2024.36.
W.K. is supported by the National Research Foundation of Korea (NRF) grant funded by the Korea government (MSIT) (RS-2024-00342488). 
The findings of this research are drawn from observations conducted using POL-2 instrument of James Clerk Maxwell Telescope (JCMT). JCMT is operated by the East Asian Observatory on behalf of The National Astronomical Observatory of Japan; Academia Sinica Institute of Astronomy and Astrophysics; the Korea Astronomy and Space Science Institute; the National Astronomical Research Institute of Thailand; Center for Astronomical Mega-Science (as well as the National Key R\&D Program of China with No. 2017YFA0402700). Additional funding support is provided by the Science and Technology Facilities Council of the United Kingdom and participating universities and organizations in the United Kingdom and Canada.
\facilities{James Clerk Maxwell Telescope (JCMT), Herschel Space Observatory.}
\software{Astropy \citep{2013A&A...558A..33A,2018AJ....156..123A,2022ApJ...935..167A}, Aplpy \citep{2019zndo...2567476R}.}
\end{acknowledgments}

\appendix

\section{Dependence of $B$-field Strength on Gas Volume Density}   \label{apdx:B_n(H2)}
\begin{figure}
    \centering   
    \includegraphics[width=0.6\columnwidth]{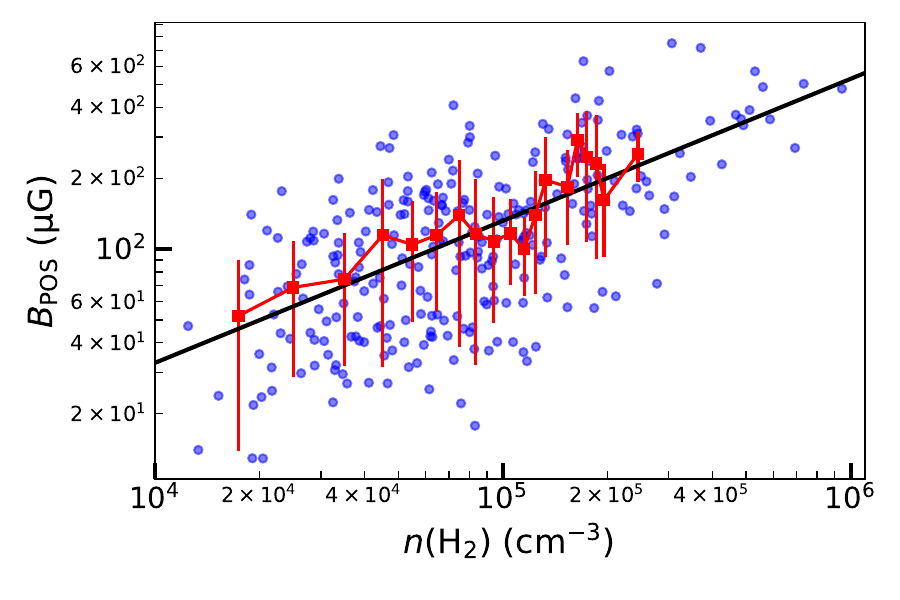}
 \caption{Variation of $B_{\rm POS}$ with $n({\rm H}_{2})$. The red curve shows the binned mean, and the error bars represent the standard deviation within each bin. The black line corresponds to the best-fit power law to the binned data, yielding a slope of $0.60\pm0.10$. An increase in the $B$-field strength is observed with increasing gas volume density.}
    \label{fig:B_vs_n_A2}
\end{figure}

Figure \ref{fig:B_vs_n_A2} shows the variation in the $B$-field strength derived using the \citetalias{2021A&A...647A.186S} method as a function of the volume density for Serpens Main. The $B$-field strength increases with increasing gas volume density. 
The scaling relation between $B$-field strength and density ($B \propto \rho^{k}$) determines the magnetic support of a cloud, which becomes important in understanding the formation and evolution of clouds. A power law relation between $B_{\rm POS}$ and $n({\rm H}_{2})$ is derived using the binned data and their associated uncertainties. The fit is shown by a black line in the log-log space and it yields a power law exponent of $0.60\pm0.10$. 
The value of $k > 0.5$ signifies the weak field model with flux freezing, and for a strong field model with ambipolar diffusion, $k < 0.5$ is predicted \citep{1999ASIC..540..305M, 2012ARA&A..50...29C, 2022ApJ...925...30L}. 
Here, we also note the bias toward the derived power law exponent due to the dependence $B_{\rm POS} \propto n({\rm H}_{2})^{1/2}$ of the \citetalias{2021A&A...647A.186S} method.  
The variation of $B_{\rm POS}$ with $n({\rm H}_{2})$ highlights the role of density in enhancing the $B$-field strength, and it emphasizes the interplay between these properties.

\section{Map of $B$-field Strength from the DCF Method}   \label{apdx:B_dcf}
\begin{figure}
    \centering
    \includegraphics[width=0.8\textwidth]{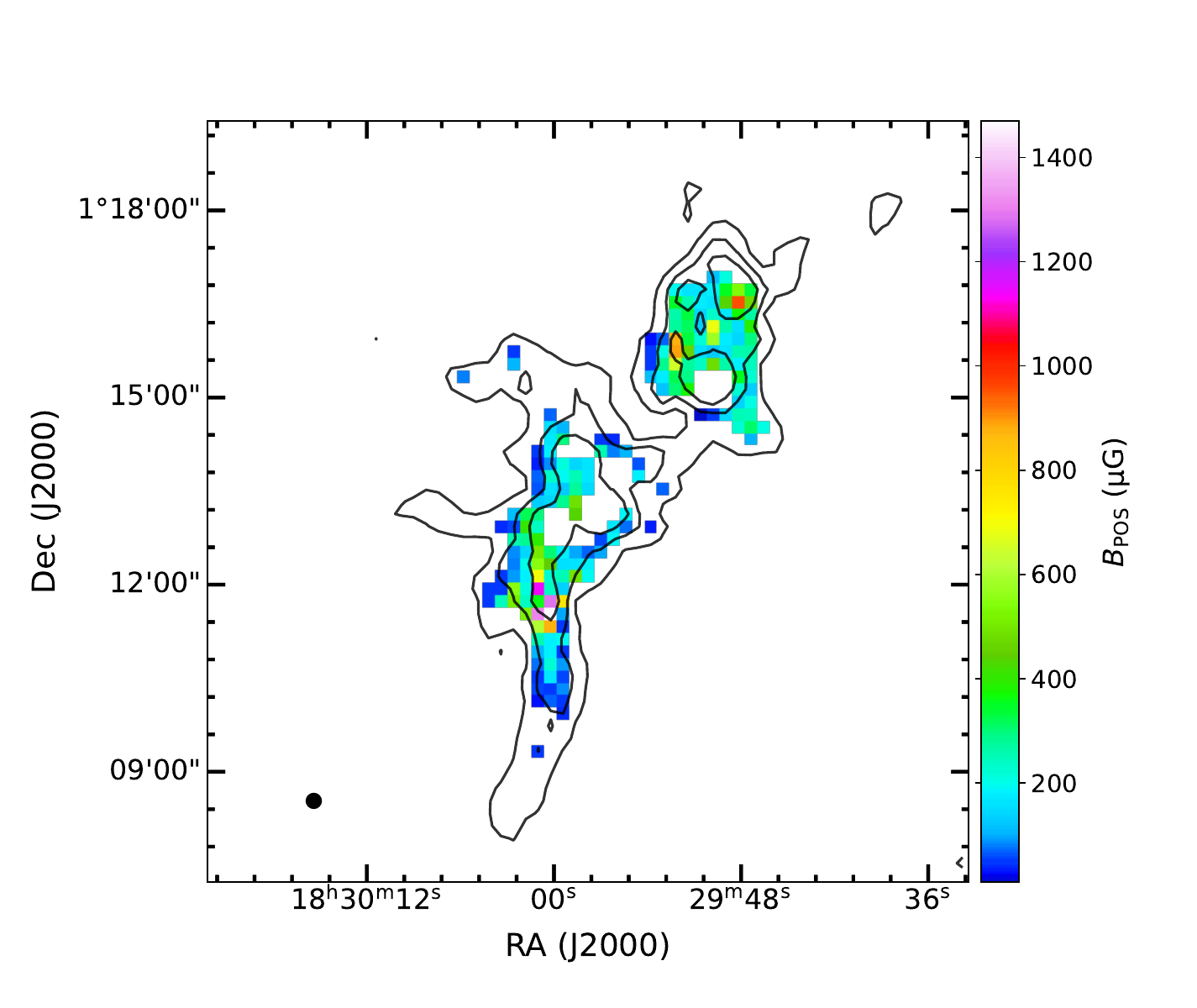}
    \caption{The map of $B_{\rm POS}$ strength using the DCF method. The contours represent the `$I$’ levels at 0.05, 0.2, and 0.5 Jy/beam.}
    \label{fig:dcf_B}
\end{figure}

We also derived the $B_{\rm POS}$ strength using the DCF method (Equation \ref{eq:b_dcf1}). 
Here, we restrict the analysis to regions where $\sigma_{\rm \theta} <$ 25\degr\, as Equation \ref{eq:b_dcf1} yields reliable results only within this range \citep{2004Ap&SS.292..225C, 2001ApJ...546..980O}.
The derived map of $B_{\rm POS}$ using the DCF method is shown in Figure \ref{fig:dcf_B}.  
The magnetic field strength exceeds 800 $\mu$G in a few pixels, while the majority of values remain below 400 $\mu$G. The field strength increases toward denser regions. A relatively higher value of $B_{\rm POS}$ is obtained using the DCF method in comparison with the \citetalias{2021A&A...647A.186S} method.  
\citet{2024ApJ...977...32C} applied both the \citetalias{2021A&A...647A.186S} and DCF methods to estimate the magnetic field strength in cores and filamentary structures in IC 348. Their results show that the \citetalias{2021A&A...647A.186S} method yields systematically lower magnetic field strengths compared to the DCF method. Similarly, \citet{2024ApJ...967..157L} used both techniques to construct a pixel-by-pixel map of the magnetic field strength of G28.37+0.07 and found relatively lower values from the \citetalias{2021A&A...647A.186S} method. A comparable trend was reported by \citet{2024A&A...684A.212B} in their study of NGC 2024, where the \citetalias{2021A&A...647A.186S} method produced lower magnetic field estimates than the DCF method.
Since both methods depend on the parameter $\sigma_{\rm \theta}$, with all other parameters being identical, we find that the largest discrepancy in the $B$-field strength between the DCF and  \citetalias{2021A&A...647A.186S} methods occurs at lower values of $\sigma_{\rm \theta}$.


\bibliography{ref}{}
\bibliographystyle{aasjournalv7}

\end{document}